\documentclass[
eqsecnum,graphics,floatfix,
nofootinbib,notitlepage,superscriptaddress,
nobibnotes,aps,prd,longbibliography]{revtex4-1}

\global\arraycolsep=2pt
\usepackage{amsmath}
\usepackage{amssymb}
\usepackage{graphicx}
\usepackage{bm}
\usepackage{xcolor}
\usepackage[normalem]{ulem} 

\DeclareMathOperator{\Tr}{Tr}

\newcommand{\psumbar}{\sum\nolimits^\prime}
\DeclareMathOperator*{\psum}{\psumbar}
\DeclareMathOperator{\tr}{tr}
\newcommand{\TE}{\text{TE}}
\newcommand{\TM}{\text{TM}}

\newcommand{\const}{\text{const}}

\newcommand{\be}{\begin{equation}}
\newcommand{\ee}{\end{equation}}
\newcommand{\bea}{\begin{eqnarray}}
\newcommand{\eea}{\end{eqnarray}}

\begin{document}
\title{Casimir Self-Entropy of Nanoparticles with Classical Polarizabilities:
Electromagnetic Field Fluctuations}

\author{Yang Li}
\email{leon@ncu.edu.cn}
  \affiliation{Department of Physics, Nanchang University, Nanchang 330031, China}
  \affiliation{Institute of Space Science and Technology, Nanchang University, Nanchang 
330031, China}
  \author{Kimball A. Milton}
  \email{kmilton@ou.edu}
  \affiliation{H. L. Dodge Department of Physics and Astronomy, University of Oklahoma, 
Norman, OK 73019, USA}
  \author{Prachi Parashar}
  \email{Prachi.Parashar@jalc.edu}
  \affiliation{John A. Logan College, Carterville, IL 62918, USA}
  \author{Gerard Kennedy}
  \email{g.kennedy@soton.ac.uk}
  \affiliation{School of Mathematical Sciences, University of Southampton, Southampton,
  SO17 1BJ, UK}
  \author{Nima Pourtolami}
  \email{nima.pourtolami@nbc.ca}
  \affiliation{National Bank of Canada, Montreal, QC H3B 4S9, Canada}
  \author{Xin Guo}
  \email{guoxinmike@ou.edu}
  \affiliation{H. L. Dodge Department of Physics and Astronomy, University of Oklahoma, Norman, OK 73019, USA}

\date{\today}
\begin{abstract}
    Not only are Casimir interaction entropies not guaranteed to be positive, but also,
    more strikingly, Casimir self-entropies of bodies can be negative.
    Here, we attempt to interpret the physical origin and meaning of these negative 
self-entropies by investigating the Casimir self-entropy of a neutral spherical
    nanoparticle. After extracting the polarizabilities of such a particle by examining
    the asymptotic behavior of the scattering Green's function, we compute the
    corresponding free energy and entropy. Two models for the nanoparticle,  
namely a spherical
    plasma $\delta$-function shell and a homogeneous dielectric/diamagnetic ball, are
    considered at low temperature, because that is
    all that can be revealed from a nanoparticle perspective.
    The second model includes a 
    contribution to the entropy from the
    bulk free energy, referring to the situation where the medium inside or outside
    the ball fills all space, which must be subtracted on physical grounds in order to
    maintain consistency with van der Waals interactions, 
    corresponding to the self-entropy of each bulk.
    (The van der Waals calculation is described in
 Appendix A.)
    The entropies so calculated agree with known results in the low-temperature limit, 
    appropriate for a small particle, and are negative. 
    But we suggest that the negative self-entropy is simply 
    an interaction entropy, the difference between the total entropy and
    the blackbody entropy of the two    bulks,   outside or inside of the
    nanosphere. The vacuum entropy is always positive and overwhelms
    the interaction entropy.  Thus
 the interaction entropy can be negative, without contradicting
    the principles of statistical thermodynamics. Given
    the intrinsic electrical properties of the nanoparticle,
    the self-entropy arises from its interaction with the thermal vacuum permeating 
all space.     Because
    the entropy of blackbody radiation by itself plays an important role, 
    it is also discussed, including dispersive
    effects, in detail. 
\end{abstract}
\maketitle

\section{Introduction}
\label{I}
\par When an electromagnetic quantum field coexists in thermal equilibrium with bodies or
boundaries, the entropy of the system will be altered by 
their presence, which may lead to nontrivial phenomena implying novel 
physics and applications. This additional entropy, known as Casimir entropy, was 
recognized as a physical quantity in the debate on how
to model a metal within a finite-temperature environment, when evaluating the Casimir 
force between metal plates.  Initially, 
there were claims that the Drude model resulted in
Casimir entropies inconsistent with the third law of 
thermodynamics~\cite{bezerra2004violation,%
klimchitskaya2015casimir,korikov2016casimir,klimchitskaya2017low}, but it has been shown 
that this does not occur for real materials~\cite{brevik2006thermal,milton2012thermal}.
Experimentally, results favoring both the plasma 
model~\cite{Decca2005Precise,Banishev2013Casimir,Bimonte2016Isoelectronic,Liu2019Examining,klimchitskaya2021},  which does not include dissipation, and the Drude
model~\cite{Sushkov2011Observation,Garcia2012Casimir}, which does,
have been reported. This inconsistency 
with the physically motivated Drude model
has not yet been resolved, but the subject of Casimir entropy by itself has drawn much 
attention.

\par Casimir interaction entropy, caused by the interactions between two or more bodies
via the fluctuating quantum electromagnetic field, has been intensively investigated for many years due to its
fascinating properties, such as its negativity~\cite{bezerra2002thermodynamical,%
bezerra2002correlation, canaguier2010thermal,rodriguez2011casimir,ingold2015geometric,%
milton2015negative,Disentangling2015Umrath,milton2017negative}. Less investigated,
however, is the Casimir self-entropy, resulting from the self-interaction of a single 
body, which provides us with further intriguing possibilities and puzzles to be 
understood.
We will concentrate on a new approach to this problem in this paper.

\par In Ref.~\cite{Li2016Casimir}, we evaluated the Casimir self-entropy of a plasma
$\delta$-function plate (PDP), with the aim of justifying the widely accepted hypothesis 
that the
negative Casimir interaction entropy would always be compensated by corresponding 
positive self-entropies. We obtained analytic formulas for the transverse electric (TE) 
and transvese magnetic (TM) contributions to the PDP self-entropy. 
They both satisfy the third law of
thermodynamics, in that the entropy vanishes as the temperature goes to zero, and indeed 
the total
self-entropy is positive, although the TE contribution is always negative. But, in the 
strong-coupling limit, which is the perfectly conducting (PC) case, the total 
self-entropy approaches zero, which eliminates the possibility that it can cancel the 
negative interaction entropy between a PC sphere and a PC plate. In 
Refs.~\cite{Li2016Casimir,milton2017negative}, we showed, however, that the Casimir 
self-entropy of a PC sphere precisely cancels the most negative part of the interaction 
entropy between a sphere and a plate. Then we generalized our study of Casimir 
self-entropy to the model of a plasma $\delta$-function spherical shell (PDS). Various 
regularization schemes were employed to evaluate the TE and TM self-entropies of a PDS 
in limiting cases in Ref.~\cite{Milton2017CasimirSelf}. It was especially surprising 
to find that, when the coupling was weak enough, both the TE and TM self-entropies are 
negative---See Eq.~(\ref{weaklambda}). Bordag and Kirsten examined the same plasma-shell
model~\cite{bordag2018free,bordag2018entropy}, but obtained
 somewhat different results; their results were technically equivalent
 to those in Ref.~\cite{Milton2017CasimirSelf}, differing only in certain subtractions. 
For more  detailed comparisons, please refer to
Ref.~\cite{milton2019remarks}. Most recently, we utilized a numerical method, based on 
the Abel-Plana formula, to elucidate general behaviors of PDS 
self-entropies~\cite{Li2021Negativity}, which confirms the results in
Ref.~\cite{Milton2017CasimirSelf} and clearly demonstrates the existence of negative
self-entropy. So, in contrast to the naive hypothesis, Casimir self-entropy can be 
quite nontrivial, and its negativity needs to be better understood.

\par According to Ref.~\cite{Milton2017CasimirSelf}, the leading terms of TE and TM PDS
self-entropies are negative and of the first order in the coupling $\lambda_0$, 
specifically,
\begin{eqnarray}
\label{eqI.1}
S^{\TE}_{(1)}=-\lambda_0\bigg(\frac{t}{6}+\frac{1}{2 t}-\frac{1}{2}\coth t\bigg),\quad
S^{\TM}_{(1)}=-\lambda_0\bigg(\frac{t}{18}-\frac{1}{2 t}+\frac{1}{2}\coth t\bigg),
\label{weaklambda}
\end{eqnarray}
in which $t=2\pi aT$, $a$ is the radius of the spherical shell, and $T$ is the temperature.
The terms linear in $\lambda_0$ might be supposed to originate from the self-interaction 
of each point in the material, in analogy with $\lambda\phi^4$ theory, although here the 
coupling refers to the entire surface of the sphere. It is conventional wisdom that this 
kind of self-interaction should be subtracted off, as a ``tadpole'' term; doing so here, 
however, would destroy the passage to the perfectly-conducting limit.
The appearance of negative Casimir self-entropy and its linear dependence on the coupling 
in PDS imply the necessity of investigating the influence of the self-interaction more 
delicately, so that the meaning of negative Casimir self-entropy and the remaining 
divergences encountered in spherical systems, or even of their consequences in reality, 
could be clarified.

For a homogeneous dielectric ball (DB), it is known that the bulk contributions must 
be subtracted \cite{milton2020self}, but even with this so-called bulk subtraction, the 
zero-temperature calculations are plagued with ambiguous 
divergences~\cite{milton2018casimir}, except for special cases where the speed of light 
is the same inside and outside the spherical 
boundary~\cite{milton1978casimir,brevik1982casimir}. Recently, Avni and 
Leonhardt~\cite{avni2018casimir}
claimed that their subtracted physical Casimir stresses on a dielectric ball yields an 
energy, depending linearly on the susceptibility in the weak-coupling limit, 
and thus violating the interpretation in terms of van der Waals forces. Some of us 
have argued that their conclusions are erroneous~\cite{milton2020self}. This 
matter is still being disputed \cite{leonhardt2021}. Here, we will extend
the calculations of Ref.~\cite{milton2020self} to finite temperature.

\par We  focus on a system composed of a
nanoparticle that interacts with a thermal electromagnetic background (blackbody
radiation) and that is characterized by macroscopic polarizability parameters, 
inferred from the
large-distance behavior of the Green's functions describing
electromagnetic scattering. As it is viewed from
far away, the particle appears as a point.
The entropy can be calculated directly in terms of these polarizabilities in a standard 
way. In this manner, we hope to shed light on the self-interaction influences alone. In  
Sec.~\ref{TC}, we extract the 
polarizabilites, expressed in terms of the reflection coefficients in the scattering 
Green's function. We consider two models for the nanoparticle,
the PDS model mentioned above, and the homogeneous nondispersive dielectric/diamagnetic 
ball. We compute the corresponding entropies in Sec.~\ref{PE}. The 
contributions depending linearly on the polarizability  are clearly identified. In 
Sec.~\ref{BFE}, we show the necessity, in the DB case, for subtracting the bulk 
contributions to achieve consistency with known results, and with the interpretation in 
terms of the van
der Waals interactions between the constituents of the nanoparticle. 
This bulk subtraction corresponds to removal of the entropy
change due to the replacement of a volume of vacuum by 
a corresponding volume of dielectric material.  In Appendix \ref{Appb}
we show that negative self-entropies correspond to what we 
call interaction entropies, which can be of any
sign, and are overwhelmed by the positive blackbody entropy.
Because the interaction with the blackbody radiation field plays an 
important role, in Sec.~\ref{vac},
we examine the entropy of the background, with and without permittivity/permeability,
including dispersion described by the
plasma model. The blackbody entropy is well-known in vacuum, but less so 
in a homogeneous medium. Especially interesting is the behavior of the entropy in 
the presence of a dispersive background. Conclusions are presented in Sec.~\ref{C}.
Appendix \ref{appa} provides evidence that the bulk subtraction for the
dielectric sphere is required for all temperatures. Appendix \ref{appa} also supplies a 
derivation of the low-temperature free energy of a dilute dielectric sphere, based
on summation of van der Waals interactions, applying a variation of a method used two 
decades ago by Barton \cite{bartoniv}.  The results for the entropy agree with those 
found in the main text. The results are also confirmed by the analysis
presented in Appendix \ref{Appb}. Appendix \ref{appc} shows that our results for the
low-temperature 
self-entropy for a dielectric/diamagnetic sphere follow immediately by extending
the zero-temperature self-energy derived long ago 
\cite{milton1980,milton1997casimir,miltonbook} to finite temperature.
Natural units $\hbar=\varepsilon_0=\mu_0=c=k_B=1$ are used throughout.
That is, we use rationalized Heaviside-Lorentz units, where the relation
between Gaussian and Heaviside-Lorentz polarizabilities is given by $\alpha^{\rm HL}
=4\pi \alpha^{\rm G}$.

\section{Extraction of Classical Polarizabilities}
\label{TC}

\par In this paper, we work with macroscopic electromagnetic theory, written in terms of 
Euclidean frequencies. Firstly, to see that bulk materials and point-like particles could 
be dealt with on the same footing, here we show how a small spherical particle of 
material behaves as a microscopically large but macroscopically small object, which we 
will refer to as a  ``nanoparticle'' for short.
\subsection{Classical Polarizabilities}
\label{TC.CP}

The Green's dyadic, $\bm{\Gamma}$, for
a given Euclidean frequency, $\zeta$, satisfies
\begin{eqnarray}
\label{eqTC.CP.1}
\left[
-\bm{\varepsilon}(\zeta,\mathbf{r})
-\frac{\bm{\nabla}\times\bm{\mu}^{-1}(\zeta,\mathbf{r})\cdot\bm{\nabla}\times\bm{1}}
{\zeta^2}\right]\cdot\bm{\Gamma}(\zeta;\mathbf{r},\mathbf{r}')
=\bm{1}\delta(\mathbf{r}-\mathbf{r}').
\end{eqnarray}
Without losing much generality, suppose the permittivity and permeability of the system
are both isotropic and inhomogeneous only in the radial direction.   For points outside 
the object,
the Green's dyadic, $\bm{\Gamma}(\zeta;\mathbf{r},\mathbf{r}')$, is written simply as
\cite{milton1997casimir}
\begin{subequations}
\label{eqTC.CP.2}
\begin{eqnarray}
\label{eqTC.CP.2a}
\bm{\Gamma}(\zeta;\mathbf{r},\mathbf{r}')=\sum_{l=1}^{\infty}\sum_{m=-l}^{l}
\left[-\bm{\nabla}\times
g_{\zeta,l}^H(r,r')\mathbf{X}^m_l(\Omega)\mathbf{X}^{m*}_l(\Omega')\times\overleftarrow{
\bm{\nabla}}'-\zeta^2g_{\zeta,l}^E(r,r')\mathbf{X}^m_l(\Omega)\mathbf{X}^{m*}_l(\Omega')
\right],
\end{eqnarray}
in which the vector spherical harmonics are defined as
\begin{eqnarray}
\label{eqTC.CP.2b}
\mathbf{X}^m_l(\Omega)=\frac{\mathbf{L} Y^m_l(\Omega)}{\sqrt{l(l+1)}},\quad
\mathbf{L}=\mathbf{r}\times\frac1i \bm{\nabla}.
\end{eqnarray}
The reduced Green's functions $g^E_{\zeta,l},\ g^H_{\zeta,l}$ satisfy the equations
\begin{eqnarray}
\label{eqTC.CP.2c}
\left[-r\frac{d}{dr}\frac{1}{(\mu,\varepsilon)}\frac{d}{dr}r+\frac{l(l+1)}
{(\mu,\varepsilon)}
+(\varepsilon,\mu)\zeta^2r^2\right]g^{(E,H)}_{\zeta,l}(r,r')=\delta(r-r').
\end{eqnarray}
\end{subequations}
The solutions for the TE and TM Green's functions in vacuum outside a spherical particle 
of radius $a$ are
($\kappa=|\zeta|$)
\label{eqTC.CP.3}
\be
g_{\zeta,l}^{E,H}(r,r')=\frac{1}{\kappa rr'}\left[s_l(\kappa r_<)e_l(\kappa r_>)
+R_l^{E,H}(\kappa a)e_l(\kappa r)e_l(\kappa r')\right], \quad r,r'>a,
\ee
in terms of the appropriate reflection coefficients (Mie coefficients) for the particle,
where $s_l$ and $e_l$ are modified Riccati-Bessel functions.
Explicit examples will be given in the following. We refer to the terms in the 
Green's function
proportional to the reflection coefficients as the scattering parts.

Imagine that source and field points are far from the particle, $r,r'\gg a$.
Then, the
reflection coefficients are to be evaluated for small values of $\kappa a$. In general, 
this means that only the $l=1$ term in the scattering Green's dyadic needs to be 
retained, since $R_1\gg R_{l\ne1}$ in this limit. This follows from the behavior for 
small arguments of the modified Riccati-Bessel functions,
\begin{eqnarray}
\label{eqTC.E.DS.3}
e_l(x)\sim \frac1{x^l}\frac{2^l \Gamma(l+1/2)}{\sqrt{\pi}},\quad
s_l(x)\sim x^{l+1}\frac{\sqrt{\pi}}{2^{l+1} \Gamma(l+3/2)},\quad x\ll 1.
\end{eqnarray}

\par
\begin{figure}
 \centering
 \includegraphics[width=5 cm]{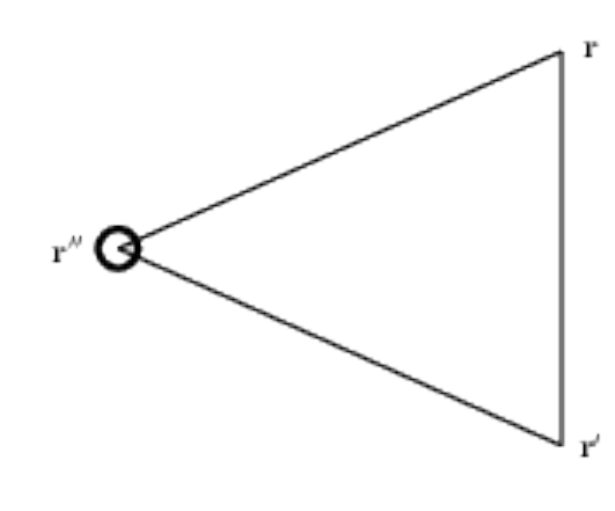}
 \caption{\label{scattfig} Sketch of the scattering process used to extract the 
polarizability of the  nanoparticle.} 
\end{figure}

 On the other hand, suppose the permittivity and permeability of the particle are 
written in terms of electric and magnetic polarizabilities, 
$\bm{\alpha}$ and $\bm{\beta}$, as $\bm{\varepsilon}(\zeta,\mathbf{r})=
\bm{1}+\bm{\alpha}(\zeta)\delta(\mathbf{r})$ and $
\bm{\mu}^{-1}(\zeta,\mathbf{r})=\bm{1}-\bm{\beta}(\zeta)\delta(\mathbf{r})$. 
Then, based on a generalization of the method given in Ref.~\cite{schwinger2000classical},
 pp.~277--8, we can identify the polarizabilities of the particle, by analyzing the above 
Green's dyadic, (\ref{eqTC.CP.2}). 
As shown in Fig.~\ref{scattfig}, the free Green's dyadic, 
$\bm{\Gamma}_0(\zeta;\mathbf{r,r'})$, is represented
 by the line going directly from $\mathbf{r'}$ to $\mathbf{r}$.
 The scattering part has a propagator going from $\mathbf{r'}$ to the nanoparticle,
 $\bm{\Gamma}_0(\zeta;\mathbf{r'',r'})$, and a second one
 going from the nanoparticle to the observer at $\mathbf{r}$, 
 $\bm{\Gamma}_0(\zeta;\mathbf{r,r''})$.  The interaction is effected via the
 polarizability of the nanoparticle, located at $\mathbf{r''=0}$.  The sum of these 
 two contributions gives the total Green's dyadic.  Because the particle
 is small, the single scattering approximation is
 sufficient. 
 The scattering part of the TE part of the Green's dyadic can be written in terms of
 $\bm{\Gamma}_0$ and the polarizabilities of the particle as
\begin{eqnarray}
\label{eqTC.CP.4}
\bm{\Gamma}_{\rm Sc}^E(\zeta;\mathbf{r},\mathbf{r}')
-\bm{\Gamma}_0(\zeta;\mathbf{r},\mathbf{r}'')\times\overleftarrow{
\bm{\nabla}}''\cdot\frac{\bm{\beta}(\zeta)}{\zeta^2}\cdot\bm{\nabla}''
\times\bm{\Gamma}(\zeta;\mathbf{r}'',\mathbf{r}')
+
\bm{\Gamma}_0(\zeta;\mathbf{r},\mathbf{r}'')
\cdot\bm{\alpha}(\zeta)\cdot\bm{\Gamma}(\zeta;\mathbf{r}'',\mathbf{r}'),\quad \mathbf{r}''
\rightarrow\bm{0},
\end{eqnarray}
where the magnetic polarizability, $\bm{\beta}$, couples to the magnetic field, given by
Faraday's law, $\bm{\nabla}\times\mathbf{E}=-\zeta \mathbf{B}$, while the electric
polarizability, $\bm{\alpha}$, couples to the electric field. ($\bm{\Gamma}$ represents
a vacuum expectation value of the product of electric fields.)
In Eq.~(\ref{eqTC.CP.4}) 
we have assumed that the particle has no extent. Schematically, the
effective polarizabilities $\bm{\alpha}$ and $\bm{\beta}$ satisfy the following relation
\begin{eqnarray}
\label{eqTC.CP.5}
\bm{\Gamma}_{\rm Sc}^E
-\bm{\Gamma}_0\left[
\times\overleftarrow{\bm{\nabla}}\frac{\bm{\beta}}{\zeta^2}\bm{\nabla}\times
-\bm{\alpha}\right]\bm{\Gamma}_0,
\end{eqnarray}
where we have noted that, because the polarizabilities are small, we may replace 
$\bm{\Gamma}$
by $\bm{\Gamma}_0$ on the right. 

Now we use the orthonormality properties of the vector spherical harmonics,
\begin{subequations}
\label{eqTC.CP.6}
\bea
\int d\Omega \,\mathbf{X}^*_{lm}(\Omega)\cdot\mathbf{X}_{l'm'}(\Omega)&=&\delta_{ll'}
\delta_{mm'},
\label{eqTC.CP.6a}
\eea
\bea
\int d\Omega[f(r')\mathbf{X}_{lm}(\Omega)]^*\cdot [\bm{\nabla}\times
g(r)\mathbf{X}_{lm}(\Omega)]&=&0,
\label{eqTC.CP.6b}
\eea
\bea
\int d\Omega \, [\bm{\nabla}'\times f(r')\mathbf{X}_{lm}(\Omega)]^*\cdot [\bm{\nabla}
\times g(r)\mathbf{X}_{lm}(\Omega)]&=&\frac1{rr'}\left[\frac{d}{dr'}(r'f(r')^*)
\frac{d}{dr}(r g(r))
+l(l+1)f(r')^*g(r)\right]\delta_{ll'}\delta_{mm'}.\quad
\label{eqTC.CP.6c}
\eea
\end{subequations}
Then, because $l=1$ dominates for a small particle, we see that only the magnetic
polarizability term contributes to the TE scattering Green's dyadic,
which we take to be isotropic:\footnote{We might anticipate that
the polarizabilities are not isotropic, but that
$\bm{\beta}=\beta(\bm{1}-\mathbf{\hat r \hat r})$.  But it
is easily checked that any radial component of
$\bm{\beta}$ does not contribute to
Eq.~(\ref{eqTC.CP.4}).}
\be
\label{eqTC.CP.7}
\beta^E=\frac32\frac{4\pi}{\kappa^3}R_1^E,\quad \alpha^E=0.
\ee
Similar arguments apply to the TM contribution. The TM part of $\bm{\Gamma}_{\rm Sc}$ is
\be
\label{eqTC.CP.8}
\bm{\Gamma}^H_{\rm Sc}(\zeta;\mathbf{r,r'})\bigg|_{l=1}=
-R^H_1\bm{\nabla}\times \frac{e_1(\kappa r)e_1(\kappa r')}{\kappa rr'}
\sum_m \mathbf{X}_{1m}(\Omega)
\mathbf{X}^*_{1m}(\Omega')\times\overleftarrow{\bm\nabla'}.
\ee
The decomposition in terms of scattering with the electric and magnetic polarizabilites of
 the particle is
\bea
\label{eqTC.CP.9}
&&
\bm{\Gamma}^H_{\rm Sc}(\zeta;\mathbf{r,r'})
\sim
\sum_{m,m'}\bigg\{
\bm{\nabla}\times\frac{e_1(\kappa r)s_1(\kappa r'')}
{\kappa rr''}\mathbf{X}_{1m}(\Omega)\mathbf{X}^*_{1m}(\Omega'')\times
\overleftarrow{\bm{\nabla}}''\cdot
\alpha^H\cdot\bm{\nabla}''
\times\frac{s_1(\kappa r'')e_1(\kappa r')}{\kappa r''r'}
\mathbf{X}_{1m'}(\Omega'')\mathbf{X}^*_{1m'}(\Omega')\times
\overleftarrow{\bm{\nabla}}'
\nonumber\\
&&
\mbox{}-\frac1{\kappa^2}\bm{\nabla}\times\frac{e_1(\kappa r)s_1(\kappa r'')}
{\kappa rr''}\mathbf{X}_{1m}(\Omega)\mathbf{X}^*_{1m}(\Omega'')
\times\overleftarrow{\bm{\nabla}}''\times\overleftarrow{\bm{\nabla}}''\cdot
\beta^H\cdot\bm{\nabla}''\times\bm{\nabla}''\times
  \frac{s_1(\kappa r'')e_1(\kappa r')}{\kappa r''r'}
\mathbf{X}_{1m'}(\Omega'')\mathbf{X}^*_{1m'}(\Omega')\times\overleftarrow{\bm
{\nabla}}'\bigg\}.\nonumber\\
\eea
Each double curl can be replaced by $-\kappa^2$, and then it is evident
that the second term above vanishes in the $r''\to 0$ limit.  Employing
the averaging over solid angles as above for the first term,
and using the identity
(\ref{eqTC.CP.6c}), we see immediately that the electric polarizability arises
from the TM scattering Green's dyadic:
\be
\label{eqTC.CP.10}
\alpha^H=\frac32\frac{4\pi}{\kappa^3}R_1^H,\quad \beta^H=0.
\ee

\subsection{Examples}
\label{TC.E}
\par To be more specific, we now turn to particular models.
\subsubsection{$\delta$-Function Spherical Shell}
\label{TC.E.DS}
\par To get a first indication of how the microscopic structure of a nanoparticle 
influences its macroscopic behavior, let us consider an example previously investigated 
in Ref.~\cite{Milton2017CasimirSelf}, namely the $\delta$-function spherical shell of 
radius $a$. Suppose the permittivity and permeability of the system are 
$\bm{\varepsilon}=\bm{1}+\lambda_ea(\bm{1}-\hat{\mathbf{r}}\hat{\mathbf{r}})\delta(r-a)$ 
and
$\bm{\mu}=\bm{1}+\lambda_ma(\bm{1}-\hat{\mathbf{r}}\hat{\mathbf{r}})\delta(r-a)$, 
respectively. According to the arguments in Ref.~\cite{parashar2017electromagnetic}, we 
have required that the polarizabilities normal to the shell surface are zero. The TE 
reflection coefficient is given by \cite{parashar2017electromagnetic,milton2011local}
\begin{subequations}
\label{eqTC.E.DS.1}
\begin{eqnarray}
\label{eqTC.E.DS.1a}
R_l^E=
\frac{
\lambda_ma\kappa s_l^{\prime2}(\kappa a)
-
\lambda_ea\kappa s_l^2(\kappa a)
}{
1-\lambda_ma\kappa e'_l(\kappa a)s'_l(\kappa a)
+\lambda_ea\kappa e_l(\kappa a)s_l(\kappa a)
+
\frac{\lambda_e\lambda_ma^2\kappa^2}{4}
},
\end{eqnarray}
and the corresponding TM coefficient, $R_l^H$, can be obtained with the substitution 
$\lambda_e\leftrightarrow\lambda_m$ as
\begin{eqnarray}
\label{eqTC.E.DS.1b}
R_l^H
=
\frac{
\lambda_ea\kappa s_l^{\prime2}(\kappa a)
-
\lambda_ma\kappa s_l^2(\kappa a)
}{
1-\lambda_ea\kappa e'_l(\kappa a)s'_l(\kappa a)
+\lambda_ma\kappa e_l(\kappa a)s_l(\kappa a)
+
\frac{\lambda_e\lambda_ma^2\kappa^2}{4}
}.
\end{eqnarray}
\end{subequations}

\par For simplicity, consider the case in which $\lambda_e\neq0,\lambda_m=0$,
so that $R^E$ and $R^H$ reduce to
\begin{eqnarray}
\label{eqTC.E.DS.2}
R_l^E
=
-\frac{\lambda_ea\kappa s_l^2(\kappa a)
}{
1+\lambda_ea\kappa e_l(\kappa a)s_l(\kappa a)
},\
R_l^H
=
\frac{
\lambda_ea\kappa s_l^{\prime2}(\kappa a)
}{
1-\lambda_ea\kappa e'_l(\kappa a)s'_l(\kappa a)
}.
\end{eqnarray}
This means that in the point approximation, $a\rightarrow0$,  where the small
argument approximations (\ref{eqTC.E.DS.3}) are applicable, the effective
polarizabilities of the nanoparticle,  dominated by $l=1$, are expressed as
\begin{eqnarray}
\label{eqTC.E.DS.4}
\beta=\beta^E=
-\frac{\lambda_e(\kappa a)^2/6
}{
1+\lambda_e(\kappa a)^2/3
}4\pi a^3
,\quad
\alpha=\alpha^H=\frac{
2\lambda_e/3
}{
1+2\lambda_e/3
}4\pi a^3.
\end{eqnarray}
Particularly, as in the model used in Refs.~\cite{Li2016Casimir,bordag2018entropy}, 
consider dispersion as given by a plasma model, i.e., $\lambda_e=\lambda_0/(\kappa a)^2$. 
Then, 
from Eqs.~(\ref{eqTC.E.DS.4}), the nonzero polarizabilities for the $\delta$-function 
sphere are
\begin{eqnarray}
\beta^E=-\frac{\lambda_0/6}{1+\lambda_0/3}4\pi a^3,\quad \alpha^H=\frac{\frac23
\frac{\lambda_0}{(\kappa a)^2}}
{1+\frac23\frac{\lambda_0}{(\kappa a)^2}}4\pi a^3.
\label{polsfordelta}
\end{eqnarray}
As expected, $\alpha^H \to 4\pi a^3$ in the strong-coupling limit, while in that limit
$\beta^E\to -\frac12 4\pi a^3$.  The electric polarizability possesses dispersion in 
general.

\par If we keep both $\lambda_e,\lambda_m$ nonzero, then, in the point approximation,
$\kappa a\to0$, $R_1^E$ and $R_1^H$ are approximated as
\begin{eqnarray}
R_1^E
\approx
\frac{
\frac{4}{9}\lambda_m(\kappa a)^{3}
-
\frac{1}{9}\lambda_e(\kappa a)^{5}
}{
1+\frac{2}{3}\lambda_m
+(\frac{1}{3}\lambda_e
+
\frac{\lambda_e\lambda_m}{4})(\kappa a)^2
},\quad
R_1^H
\approx
\frac{
\frac{4}{9}\lambda_e(\kappa a)^{3}
-
\frac{1}{9}\lambda_m(\kappa a)^{5}
}{
1
+
\frac{2}{3}\lambda_e
+
(\frac{1}{3}\lambda_m
+
\frac{\lambda_e\lambda_m}{4})(\kappa a)^2
}
,
\end{eqnarray}
which lead us to the magnetic and electric polarizabilities 
\begin{eqnarray}
\beta^E
=
\frac{
\frac{2}{3}\lambda_m
-
\frac{1}{6}\lambda_e(\kappa a)^{2}
}{
1+\frac{2}{3}\lambda_m
+(\frac{1}{3}\lambda_e
+
\frac{\lambda_e\lambda_m}{4})(\kappa a)^2
}4\pi a^3,\quad
\alpha^H
=
\frac{
\frac{2}{3}\lambda_e
-
\frac{1}{6}\lambda_m(\kappa a)^{2}
}{
1
+
\frac{2}{3}\lambda_e
+
(\frac{1}{3}\lambda_m
+
\frac{\lambda_e\lambda_m}{4})(\kappa a)^2
}4\pi a^3
.
\end{eqnarray}
We will not pursue the effects of $\lambda_m$ further here.

\subsubsection{Dielectric Ball}
\label{TC.E.DB}
\par As a second and more realistic example, we study a homogeneous dielectric ball of 
radius $a$ with nondispersive permittivity $\varepsilon$ and permeability $\mu$, immersed
in vacuum. The reflection coefficients are \cite{milton1997casimir}
\begin{eqnarray}
\label{eqTC.E.DB.1}
R_l^E
=
-\frac{
s_l(\widetilde{\kappa}a)s_l'(\kappa a)
-\sqrt{\frac{\varepsilon}{\mu}}s_l(\kappa a)s_l'(\widetilde{\kappa}a)
}{
s_l(\widetilde{\kappa}a)e_l'(\kappa a)
-\sqrt{\frac\varepsilon\mu}e_l(\kappa a)s_l'(\widetilde{\kappa}a)
},\quad
R^H_l
=
-\frac{
\sqrt{\frac{\varepsilon}{\mu}}s_l(\widetilde{\kappa}a)s_l'(\kappa a)
-s_l(\kappa a)s_l'(\widetilde{\kappa}a)
}{
\sqrt{\frac{\varepsilon}{\mu}}s_l(\widetilde{\kappa}a)e_l'(\kappa a)
-e_l(\kappa a)s_l'(\widetilde{\kappa}a)
}.
\end{eqnarray}
Here $\kappa=|\zeta|$, while $\widetilde{\kappa}=\sqrt{\varepsilon\mu}\kappa$. We proceed as above, and require the small $a$ limit, where
\begin{eqnarray}
\label{eqTC.E.DB.2}
R_1^E=\frac23\frac{\mu-1}{\mu+2}(\kappa a)^3+O\big[(\kappa a)^5\big],
\quad
R_1^H\sim \frac23\frac{\varepsilon-1}{\varepsilon+2} (\kappa a)^3
+O\big[(\kappa a)^5\big],\quad \kappa a\ll1,
\end{eqnarray}
Only the terms of order $(\kappa a)^3$ will survive, so it follows that the TE 
contribution to the magnetic polarizabilities is nonzero, if the ball is permeable,
\begin{eqnarray}
\label{eqTC.E.DB.3}
\alpha^E=0,\quad\beta^E=\frac{\mu-1}{\mu+2}4\pi a^3,
\end{eqnarray}
while the TM part yields an electric polarizability depending on the
permittivity,
\begin{eqnarray}
\label{eqTC.E.DB.4}
\alpha^H=\frac{\varepsilon-1}{\varepsilon+2}4\pi a^3,\quad \beta^H=0.
\end{eqnarray}
This electric polarizability is just that found in electrostatics 
\cite{schwinger2000classical}.
Again, in the perfectly conducting limit, the polarizabilities tend to their expected 
values,
\begin{eqnarray}
\label{eqTC.E.DB.5}
\alpha^H\to 4\pi a^3,\quad \beta^E\to -\frac12 4\pi a^3,\quad \varepsilon\to
\infty,\,\mu\to 0.
\end{eqnarray}

\par For both the $\delta$-function spherical shell and dielectric ball examples, the 
point-approximated polarizabilities are proportional to the volume of the nanoparticle, 
which is consistent with the small-polarizability assumption adopted in 
Eq.~\eqref{eqTC.CP.4}. 
Furthermore, although the constituents of the dielectric ball
interact, as explicitly demonstrated in Appendix \ref{appa},
the Clausius-Mossotti equation, which in our approximation 
incorporates those interactions, means that the polarizabilities are linearly
related to their corresponding microscopic counterparts; that is, the polarizability
of the nanoparticle is simply the sum of the polarizabilities of its microscopic
constituents.  This is discussed in detail in Appendix \ref{Appb}.

\section{Particle Description of Casimir Self-Entropy}
\label{PE}
\par Now we arrive at the main topic of this paper,
namely the Casimir self-entropy,
which is just the additional entropy of the thermal field
induced by a single object
in it, when the effects of the thermal blackbody field have been
properly removed.
As has been stated in Sec.~\ref{I}, we saw interesting contributions from 
self-interaction, which we would like to investigate with the illustrative particle 
models here.

\par Because of the different frequency dependencies, the contributions to the free energy 
can have quite different behavior for small polarizabilites. We compute the free energy 
from the sum over Matsubara frequencies ($\zeta_m=2\pi T m$)
\begin{eqnarray}
\label{eqPE.1}
F=\frac{T}2\sum_{m=-\infty}^\infty e^{i\zeta_m\tau}\Tr\ln(\bm{1}-\bm{\Gamma}_0\mathbf{V})
\approx-\frac{T}2\int (d\mathbf{r'})\sum_{m=-\infty}^\infty e^{i\zeta_m\tau}
\tr \bm{\Gamma}_0(\zeta_m;\mathbf{r,r'})\mathbf{V}(\zeta_m,\mathbf{r'})
\bigg|_{\mathbf{r=r'}+\bm{\rho}},
\end{eqnarray}
which has been regulated by point-splitting in time, $\tau$, and in space, $\bm{\rho}$. 
Here, we treat the potential of the isotropic polarizable point particle as
\begin{eqnarray}
\label{eqPE.2}
\mathbf{V}_e(\zeta_m,\mathbf{r'})=\bm{1}\alpha(\zeta_m) \delta(\mathbf{r'}),\quad 
\mathbf{V}_m(\zeta_m,\mathbf{r'})=\bm{1}\beta(\zeta_m)\delta(\mathbf{r'}).
\end{eqnarray}
The reason for retaining only the first order in the potential
is that the particle is small, not that the coupling, $\lambda$, $\varepsilon-1$, or
$\mu-1$, is weak.
The trace of the Green's dyadic is
\begin{eqnarray}
\label{eqPE.3}
\tr\bm{\Gamma}_0(\zeta;\bm{\rho},\bm{0})\big|_{\bm{\rho}\to\bm{0}}=\tr(\bm{\nabla
\nabla}-\bm{1}\nabla^2)\frac{e^{-|\zeta| r}}{4\pi r}
\bigg|_{r=\rho\to 0}=-2\zeta^2
\frac{e^{-|\zeta|\rho}}{4\pi \rho}, \quad\rho=|\bm{\rho}|\to 0,\label{trGamma}
\end{eqnarray}
which uses the scalar Green's function equation
\begin{eqnarray}
\label{eqPE.4}
\left(-\nabla^2+\zeta^2\right)\frac{e^{-|\zeta|r}}{4\pi r}=\delta(\mathbf{r}).
\end{eqnarray}
It is important that the spherical symmetry 
be respected by the regulator, in this case the distance between the two points, $\rho$.

\par In addition to dispersion, anisotropy of the particle could also play a significant 
role. Here, however, we cannot contradict the spherical symmetry requirement, but it might
 be expected that the particle could have a polarizability of the form 
$\bm{\alpha}=\alpha(\bm{1}-\mathbf{\hat r''\hat r''})$. However, it is easily seen that 
the radial-radial component of the Green's dyadic is zero, so this is without effect. 

\subsection{$\delta$-Function Spherical Shell}
\label{PE.DS}
\par We first consider the $\delta$-function spherical shell model above. Using the 
polarizability $\beta^E$ in Eq.~\eqref{polsfordelta}, if we only employ spatial 
point-splitting, that is, set $\tau=0$,  we find, as $\rho\to0$
\begin{eqnarray}
F^E= -\frac16\frac{\lambda_0}{1+\lambda_0/3}a^3
\left(\frac2{\pi\rho^4}-\frac{2\pi^3 T^4}{15}\right),\quad aT\ll 1,
\label{lowte}
\end{eqnarray}
where the restriction on $T$ emerges from the point-particle limit.
In contrast, $\alpha^H$ has nontrivial frequency dependence. For the weak-coupling TM 
contribution, there is an additional $1/\zeta^2$ factor, so the behavior is given by
\begin{eqnarray}
\label{eqPE.DE.2}
F^H=\frac23\frac{\lambda_0}{a^2}a^3\left(\frac1{\pi
\rho^2}+\frac\pi3 T^2\right), \quad \lambda_0\ll1,\,\, aT\ll 1.
\end{eqnarray}
For strong coupling, the TM contribution is the same form as the TE, except for the 
replacement of $\beta^E=-\frac12 4\pi  a^3$ by  $\alpha^H=-2\beta^E=4\pi a^3$. The 
divergent terms for both the TE and TM contributions are independent of temperature, so the
 weak-coupling, low-temperature entropies are with $t=2\pi a T$
\begin{eqnarray}
\label{eqPE.DE.3}
S^E=-\lambda_0\frac{t^3}{90}, \quad S^H=-\frac29\lambda_0 t,
\end{eqnarray}
which are exactly the results found in Ref.~\cite{Milton2017CasimirSelf}. The strong 
coupling limits are given by
\begin{eqnarray}
\label{eqPE.DE.4}
S_\infty^H=-2S_\infty^E=\frac{t^3}{15},\label{pcs}
\end{eqnarray}
which are consistent with the well-known perfectly-conducting sphere results for low 
temperature~\cite{balian1978electromagnetic}. Note that in weak coupling, the total 
self-entropy is negative, while it is positive in strong coupling.

\subsection{Dielectric/Diamagnetic Ball}
\label{PE.DB}
\par We now turn to the homogeneous dielectric/diamagnetic ball. Following the same 
procedure, we find for the free energies, assuming the absence of dispersion 
[see Eq.~\eqref{eqTC.E.DB.4} and Eq.~\eqref{eqTC.E.DB.3}],
\begin{eqnarray}
\label{eqPE.DB.1}
F^{H,E}=\frac1{4\pi}\left(\begin{array}{c}\alpha^H\\ \beta^E\end{array}\right)
\left(\frac2{\pi\rho^4}-\frac{2\pi^3 T^4}{15}\right).
\end{eqnarray}
If we use the point-splitting in time rather than in space,that is,
keep $\tau\ne0$, but set $\rho=0$,
 we encounter
\be \sum_{m=-\infty}^\infty e^{i\zeta_m \tau}\tr\bm{\Gamma}_0(\zeta_m)=
-\frac1{2\pi}\sum_{m=-\infty}^\infty e^{i\zeta_m\tau}
\zeta_m^2\left(\frac1\rho-|\zeta_m|\right),
\ee
where we have expanded Eq.~(\ref{trGamma}) for small $\rho$.
The first term here is proportional to a second derivative of a $\delta$-function in
$\tau$ [see Eq.~(\ref{2ndderdelta})], so is to be omitted, while the second is 
\be \frac{i}{2\pi}\left(\frac\partial{\partial\tau}\right)^3\left[\frac1{1-
e^{i2\pi T\tau}}-\frac1{1-e^{-i2\pi T\tau}}\right].
\ee
Carrying out the differentiation and expanding now in $\tau$, we obtain
\begin{eqnarray}
\label{eqPE.DB.2}
F^{H,E}=\frac1{4\pi}\left(\begin{array}{c}\alpha^H\\ \beta^E\end{array}\right)
\left(-\frac6{\pi\tau^4}-\frac{2\pi^3T^4}{15}\right),
\end{eqnarray}
where the $-3$ ratio in the coefficients of the divergence is expected on general
grounds \cite{torque1}. [See, for example, Eq.~(\ref{vacu}).]
Both Eqs.~(\ref{eqPE.DB.1}) and
(\ref{eqPE.DB.2}) have the same finite part,
which yields the total self-entropy ($t=2\pi a T$)
\begin{eqnarray}
S=S^H+S^E=\left(\frac{\varepsilon-1}{\varepsilon+2}+\frac{\mu-1}{\mu+2}\right)
\frac{t^3}{15}.\label{sheps}
\end{eqnarray}
This has the correct strong-coupling (perfectly-conducting) $\varepsilon\to\infty$, 
$\mu\to0$ limit in Eq.~\eqref{pcs}. Comparing Eq.~\eqref{eqPE.DE.3} and Eq.~\eqref{sheps}, 
we see how different models of the nanoparticle can lead to entirely
disparate behaviors of the self-entropy.
In particular, this entropy is positive for $\varepsilon>1$, 
$\mu>1$, although it could be of either sign if one of
these inequalities is violated.

\par For dilute constituents of the dielectric/diamagnetic ball, 
we evidently see terms depending linearly on the susceptibilities $\varepsilon-1$ and 
$\mu-1$ in Eq.~(\ref{sheps}).
This is extraordinary and seems, at first sight, inexplicable, considering well-established
understandings. We showed~\cite{milton2020self}, at zero temperature,
that the free energy should begin, in the dilute limit for a pure dielectric ball, as 
$(\varepsilon-1)^2$, which is understood as originating from the pairwise summation of van 
der Waals interactions. The free energy was also calculated many years ago by Nesterenko, 
Lambiase, and Scarpetta~ \cite{nesterenko2001casimir}, and by 
Barton~\cite{bartoniv,barton2001perturbative}\footnote{Barton gets an extra term,
besides the two displayed in Eq.~(\ref{eqPE.DB.2}), proportional to the area of
the sphere: $\Delta F_B=-\frac14(\varepsilon-1)^2 \zeta(3) a^2T^3$.  This discrepancy
seems not to have been resolved. We rederive this result, without
this discrepant term, by a variation of Barton's method in Appendix \ref{appa}.}
\begin{eqnarray}
\label{eqPE.DB.3}
F=\frac{23}{1536}\frac{(\varepsilon-1)^2}{\pi a}+\frac7{270}(\varepsilon-1)^2
\pi^3a^3 T^4.\label{dbfe}
\end{eqnarray}
The first term, corresponding to zero temperature, was first calculated by Milton and Ng 
(by summing van der Waals interactions)~\cite{milton1998vdw} and by Brevik, Marashevsky, 
and Milton (by expanding the Casimir energy)~\cite{brevik1999identity}. The authors of 
Ref.~\cite{nesterenko2001casimir} seem not to remark that the corresponding entropy is 
negative,
\begin{eqnarray}
S=-\frac7{540}(\varepsilon-1)^2 t^3,\quad |\varepsilon-1|\ll1.\label{nlsentropy}
\end{eqnarray}
So, the initial linear behavior in $\varepsilon-1$ seen in Eq.~(\ref{sheps}) is not 
present. Although there might be some differences between Casimir entropies of a 
nanoparticle and a bulk dielectric ball, it is still puzzling to see this discrepancy,
which will be dealt with in the following section.

\section{Bulk free energy}
\label{BFE}

Hitherto, we have been considering a point particle.  The conundrum mentioned at the
end of the preceding section arises when 
we recognize that an extended object appears as a point far away from the object.  
The formula obtained for the additional free energy resulting from the insertion of the 
particle into the thermal bath therefore includes everything. However, when one looks at 
the ball in the near field and considers it as an extended object, one must recognize
that an additional part of the free energy comes from the replacement 
of a point particle 
(of zero volume) by a medium of finite volume---in effect, this finite volume of the 
medium “displaces” what was previously considered the same volume of vacuum in 
the far field, point particle, perspective. This part of the additional free energy must 
therefore be subtracted in order to be left with the true additional free energy due to 
the interaction, and this subtraction is the finite temperature bulk subtraction. 
The need for this subtraction here is simply because the initial setup regarded the 
particle as a point, and the finite extension of the particle itself contributes a change
to the free energy of that volume, which is in a sense extraneous to what is being 
sought here, and so must be subtracted.

This way of extracting a meaningful self-free energy for the dielectric ball is what is
conventionally done, in order to obtain consistency with van der Waals interactions.
That is, we subtract the contribution that would be obtained if either the interior or the 
exterior medium filled all of space. This was discussed recently in detail in 
Ref.~\cite{milton2020self}, but only at zero temperature. We can follow the method 
articulated in Appendix~A of that reference. The most unambiguous way to proceed is to 
start with the pressure on the sphere, which is the discontinuity, across the surface, of 
the radial-radial stress tensor component,
\begin{eqnarray}
\label{eqBFE.1}
p^{(0)}(\varepsilon,\mu;\varepsilon',\mu';a)=T^{(0)}_{rr}(\varepsilon,\mu;a_-)
-T^{(0)}_{rr}(\varepsilon',\mu';a_+),
\end{eqnarray}
where the two stress tensors refer to a homogeneous medium, either $\varepsilon$, $\mu$, 
or $\varepsilon'$, $\mu'$ filling all space, and $a_\pm$ means the corresponding stress 
tensor is evaluated just outside or just inside the spherical boundary of the 
dielectric/diamagnetic sphere. (In our case, $\varepsilon'$ and $\mu'$ are both set equal 
to unity.) Here, in each region, using the Matsubara frequency decomposition at finite 
temperature $T$, the use of which is equivalent to that of the fluctuation-dissipation 
theorem)\footnote{Note that the formula for the stress tensor in Ref.~\cite{milton2020self}
 is inccorect by a minus sign.}
\begin{eqnarray}
\label{eqBFE.2}
T_{rr}^{(0)}(\varepsilon,\mu,a_-)=\frac1{2a^4}aT\sum_{m=-\infty}^\infty
e^{i\zeta_m\tau}\sum_{l=1}^\infty \frac{2l+1}{4\pi}f_l(x_m),
\end{eqnarray}
with  $x_m=2\pi|m|a T \sqrt{\varepsilon\mu}$, with the summand being
\begin{eqnarray}
\label{eqBFE.3}
f_l(x)=2x[s_l'(x)e_l'(x)-s_l''(x)e_l(x)]=2\frac\partial{\partial r}
\left(\frac\partial{\partial s}-\frac\partial{\partial r}\right)
\frac{s_l(rx)e_l(sx)}{x}\bigg|_{s>r>1, s\to 1}.
\end{eqnarray}
Now using the addition theorem, with $s>r$,
\begin{eqnarray}
\label{eqBFE.4}
\sum_{l=0}^\infty (2l+1)s_l(rx)e_l(sx)=\frac{xrs}{s-r}e^{-x(s-r)},
\end{eqnarray}
we see that the bulk stress tensor is
\begin{eqnarray}
\label{eqBFE.5}
T_{rr}^{(0)}(\varepsilon,\mu;a_-)
&=&
\frac{T}{4\pi a^3}\sum_{m=-\infty}^\infty
e^{i\zeta_m\tau}\frac\partial{\partial r}\left(\frac\partial{\partial s}-
\frac\partial{\partial r}\right)
\left(\frac{rs}{s-r}-\frac1{2x_m}\right)
e^{-x_m(s-r)}\bigg|_{s>r>1,s\to 1}
\nonumber\\
&=&
\frac{T}{4\pi a^3}\frac\partial{\partial r}\left(\frac\partial{\partial s}-
\frac\partial{\partial r}\right)
\bigg\{\frac{rs}{s-r}\left[\frac1{1-e^{-x
(s-r)+i\hat\tau}}+\frac{1}{1-e^{-x(s-r)-i\hat\tau}}-1\right]
\nonumber\\
&&
\qquad\qquad\qquad\mbox{}+\frac{1}{2x}
\ln\bigg[\left(1-e^{-x(s-r)+i\hat\tau}\right)\left(1-e^{-x(s-r)-i\hat\tau}\right)\bigg]
\bigg\}\bigg|_{s>r>1,s\to 1}.
\end{eqnarray}
Here, we have abbreviated $x=x_1=2\pi a T\sqrt{\varepsilon\mu}$ and $\hat\tau=2\pi T\tau$. 
When the differentiations, and the $s\to 1$ limit is carried out, and the result is 
expanded for small temporal cutoff $\tau$, we find
\begin{eqnarray}
T_{rr}^{(0)}(\varepsilon,\mu;a_-)
=\frac1{\pi^2}(\varepsilon\mu)^{3/2}\frac1{\tau^4}
+\frac{\pi^2}{45}(\varepsilon\mu)^{3/2}T^4.\label{trr}
\end{eqnarray}
The corresponding free energy is determined from the principle of virtual work:
\begin{eqnarray}
\label{eqBFE.7}
p^{(0)}(\varepsilon,\mu;1,1;a)=
-\frac1{4\pi a^2}\frac\partial{\partial a}F^{(0)}(T,a),
\end{eqnarray}
where the bulk free energy is
\begin{eqnarray}
\label{eqBFE.8}
F^{(0)}(T,a)=-\frac{4 a^3}3\left(\frac{1}{\pi\tau^4}+\frac{\pi^3T^4}{45}
\right)\left[(\varepsilon\mu)^{3/2}-1\right].
\end{eqnarray}
The first term here is the expected quartic divergence seen in Ref.~\cite{milton2020self}
(apart from the erroneous sign there),
while the second gives an entropy\footnote{\label{vacfn}It might be noted
that the entropy of the thermal vacuum (blackbody radiation) confined to the volume of
the nanosphere is $S^a_{\rm vac}=2 t^3/135$. See Eq.~(\ref{fands}); the
corresponding entropy for  the dielectric medium is given in Eq.~(\ref{semvac}).}
\begin{eqnarray}
\label{eqBFE.9}
S^{(0)}=\frac2{135}t^3 \left[(\varepsilon\mu)^{3/2}-1\right].
\end{eqnarray}
As stated at the beginning of this section, this is just the
entropy change due to the replacement of the vacuum by the dielectric/diamagnetic
medium in the volume enclosed by the spherical boundary of the nanoparticle.
When this (for $\mu=1$) is subtracted from the entropy computed in Eq.~(\ref{sheps}), and 
the result expanded in powers of $\varepsilon-1$, it is seen that the linear terms cancel 
(this occurs for the divergent contributions to the free energy as well in
\eqref{eqPE.DB.2}), and the quadratic terms combine to
\begin{eqnarray}
\label{eqBFE.10}
S^H-S^{(0)}=-\frac7{540}(\varepsilon-1)^2 t^3,\quad |\varepsilon-1|\ll1,
\end{eqnarray}
exactly the result (\ref{nlsentropy}) found in 
Refs.~\cite{nesterenko2001casimir,bartoniv,barton2001perturbative}.

We thus recognize that the deviations from the scattering contribution \eqref{sheps}
stem from the subtraction of
the bulk contribution. Intuitively, it makes sense that the linear $\varepsilon-1$ terms
in Eq.~\eqref{sheps} and Eq.~\eqref{eqBFE.9} are the same, as they arise from the
self-interaction of the medium, which has always to be subtracted to obtain a physically
measurable quantity, such as Casimir-Lifshitz force, as was recognized by Lifshitz and
co-workers in mid-1950s \cite{dlp} at arbitrary temperatures. 

\par There is no doubt that at zero temperature, the bulk Casimir energy of the medium is
divergent and should be properly ``renormalized,'' or at least subtracted, to extract
physics. But perhaps a system with a particular geometry at finite temperature provides
us with a chance to unveil the physics hiding in the nontriviality of the
divergent bulk Casimir energy. It is a necessary condition for a quantity to be
considered physical that this quantity should be unchanged no matter which regularization
scheme is employed. As shown above, Eq.~\eqref{trr} is obtained by carrying out the
$s\rightarrow1$ limit first and then keeping the leading orders of the temporal cutoff.
Alternatively, we could change the order of limits, i.e.,  take $\tau\to0$ first,
then set $r=1$, and finally seek the asymptotic behavior as $s$ goes to 1,
With this approach, the stress $T_{rr}^{(0)}$ takes the form
\begin{equation}
\label{eqBFE.11}
T_{rr}^{(0)}(\varepsilon,\mu;a_-)
=-\frac{1}{\pi^2a^4\sqrt{\varepsilon\mu}}
\bigg[
\frac{3}{(s-1)^4}+\frac{4}{(s-1)^3}+\frac{1}{(s-1)^2}
\bigg]
+
\frac{\pi^2}{45}(\varepsilon\mu)^{3/2}T^4.
\end{equation}
This spatial point-splitting yields a different divergence structure, but the 
temperature-dependent term is just the same as in Eq.~\eqref{trr},  which gives us some 
confidence in that result.
(A still different divergence structure emerges if, for example, we take the limit
$r\to s(1-\epsilon)$, $\epsilon\to0$, and then set $s=1$, but the temperature
dependence remains unchanged.)

For further discussion of the meaning of the bulk subtraction, the interaction entropy,
and the sign of the latter, see Appendix \ref{Appb}.

\section{Blackbody entropy}
\label{vac}
We now turn to the entropy of the background with which the nanoparticle
interacts.
\subsection{Vacuum entropy}
This initial discussion follows that in Ref.~\cite{Li2016Casimir}.  For a
more complete discussion, see Sec.~\ref{plasma}.
We start with the free scalar Green's function in empty space, at temperature
$T$:
\be
G(\tau,\rho)=\frac{T}{4\pi \rho}\sum_{m=-\infty}^\infty e^{i\zeta_m\tau}
e^{-|\zeta_m|\rho},\label{GT0}
\ee
in terms of the Euclidean time difference $\tau$ and the spatial separation
$\rho$, which we will regard as temporal and spatial regulators, tending to
zero.  The Matsubara sum is immediately carried out:
\be
G(\tau,\rho)=\frac{T}{4\pi R}\left(-1+\frac1{1-e^{-2\pi T(\rho-i\tau)}}+
\frac1{1-e^{-2\pi T(\rho+i\tau)}}\right).\label{GT}
\ee
To find the energy density, we apply a differential operator:
\be
u=T^{00}=\frac12(\partial^0\partial^{0\prime}+\bm{\nabla\cdot\nabla}')G(\tau,
\rho)
=\frac12\left(\frac{\partial^2}{\partial\tau^2}-\frac{\partial^2}{\partial\rho^2}
-\frac2\rho \frac{\partial}{\partial\rho}\right)G(\tau,\rho),
\ee
where we have used translational invariance and noted that $t-t'=i\tau$.
We can further replace the radial Laplacian by a second $\tau$ derivative,
because of the differential equation satisfied by the Green's function.  Thus
we simply obtain
\be
u=\frac1{2\pi^2}\frac{3\tau^2-\rho^2}{(\tau^2+\rho^2)^3}+\frac{\pi^2 T^4}{30}.
\label{vacu}
\ee
This is to be multiplied by two for electromagnetism, since the
divergenceless  Green's dyadic for electromagnetism is
\be
\bm{\Gamma}'(\tau,\rho)=(\bm{\nabla\nabla-1}\nabla^2)G(\tau,\rho).
\ee
Thus the electromagnetic free energy density and entropy density are
($f=u-Ts$).
\be
f_{\rm em}=\frac1{\pi^2}\frac{3\tau^2-\rho^2}{(\tau^2+\rho^2)^3}
-\frac{\pi^2 T^4}{45},\quad s_{\rm em}=\frac{4\pi^2 T^3}{45}. \label{fands}
\ee
The divergence structure is that found by Christensen \cite{christensen}.

\subsection{Homogeneous nondispersive background}
What happens if the vacuum is replaced by a uniform medium made of
a homogeneous dispersionless fluid characterized by permittivity $\varepsilon$
and permeability $\mu$?  Then, since the
Euclidean-frequency Green's dyadic satisfies
\be
\left(-\varepsilon-\frac1{\mu\zeta^2}\bm{\nabla\times\nabla}\times\right)\bm{
\Gamma}(\rho;\zeta)=\bm{1}\delta(\bm{\rho}),
\ee
we see that $\varepsilon\bm{\Gamma}(\rho;\zeta\sqrt{\varepsilon\mu})$ satisfies
the free Green's dyadic equation.  And since the energy density is
\be
u=\frac12(\varepsilon E^2+\mu H^2),
\ee
and the contributions from the two terms are identical, we see that
the effective scalar Green's function in space and Euclidean time is
\be
\varepsilon G(\tau,\rho)=T\sum_{m=-\infty}^\infty e^{i\zeta_m\tau}\frac{
e^{-|\zeta_m|\sqrt{\varepsilon\mu}\rho}}{4\pi\rho}.
\ee
Apart from the leading factor of T, this looks like the vacuum formula
(\ref{GT0}) with $T\to T\sqrt{\varepsilon\mu}$ and
$\tau\to \tau/\sqrt{\varepsilon\mu}$.
Thus, multiplying by 2, we obtain the vacuum energy density
\be
u_{\rm em}=-\frac1{\pi^2}\frac1{\sqrt{\varepsilon\mu}}
\frac{\rho^2-3\tau^2/\varepsilon
\mu}{(\rho^2+\tau^2/\varepsilon\mu)^3}+\frac{\pi^2}{30}(\varepsilon\mu)^{3/2}T^4.
\ee
For a purely spatial cutoff, $\tau=0$, this yields for the divergent part:
\begin{subequations}
\be
u_{\rm div}=-\frac1{\pi^2\sqrt{\varepsilon\mu}\rho^4},
\ee
while for a temporal cutoff ($\rho=0$):
\be
u_{\rm div}=\frac{3(\varepsilon\mu)^{3/2}}{\pi^2 \tau^4}.\label{div1}
\ee
\end{subequations}
In any case, the entropy density is
\be
s_{\rm em}=(\varepsilon\mu)^{3/2}\frac{4\pi^2 T^3}{45},\label{semvac}
\ee
as we already saw in Eq.~(\ref{eqBFE.9}).
[The apparent discrepancy between the divergent terms in
Eqs.~(\ref{div1}) and (\ref{eqBFE.8}) is explained at the end of this section.]

\subsection{Dispersion}
\label{plasma}
Now, suppose the background is described by a permittivity given by the plasma
model, without dissipation,
$\varepsilon=1+\omega_p^2/\zeta^2$, where $\omega_p^2=n e^2/m$ is the square of the
plasma frequency, in terms of charge carriers of charge $e$, mass $m$, and
number density $n$.  Realistically, in the universe, $n$ is very small, say
$1$--$10^{-4}$\,cm$^{-3}$ \cite{density}, so for electrons,
$\omega_p\sim 10^{-10}$--$10^{-13}$\,eV, and the number would be
much smaller if we considered hadronic matter.  The 3\,K CMB
radiation corresponds to an energy of order $10^{-4}$ eV, so $\omega_p$ is
much smaller than the lowest Matsubara frequency.

The above treatment must be improved in order to describe dispersion.
We can follow Ref.~\cite{parashar2017electromagnetic}, which says that the internal energy
of an object characterized by dispersive isotropic permittivity and permeability is
\be
U=T\sum_{m=-\infty}^\infty \Tr\left[\varepsilon+\frac12\zeta_m
\frac{d\varepsilon}{d\zeta_m}-\frac1{2\zeta_m}\bm{\nabla}\times\frac1{\mu^2}
\frac{d\mu}{d\zeta_m}\bm{\nabla}\times\right]\bm{\Gamma}.\label{generalu}
\ee
Here we consider a nonmagnetic material so the last term is not present.
Now consider a plasma model for the dispersion, $\varepsilon=                              
1+\omega_p^2/\zeta_m^2$; remarkably, then, the first two terms collapse to
\be
U=T\sum_{m=-\infty}^\infty \Tr\bm{\Gamma}.
\ee
So the only appearance of the plasma frequency is in the Green's dyadic.

The differential equation satisfied by Green's dyadic is
\be
\left(-\frac1{\zeta_m^2}\bm{\nabla\times\nabla}\times -1
-\frac{\omega_p^2}{\zeta_m^2}\right)\bm{\Gamma}(\rho;\zeta_m)=\bm{1}\delta(\bm{
\rho}).\label{gammainv}
\ee
This is solved by the following construction:
\be
\bm{\Gamma}=(1+\omega_p^2/\zeta_m^2)^{-1}\tilde{\bm \Gamma},
\ee
where, apart from a $\delta$-function term (contact term),
\be
\tilde{\bm{\Gamma}}=\tilde{\bm{\Gamma}}'-\bm{1},
\ee
the divergenceless Green's dyadic $\tilde{\bm{\Gamma}}'$ is built from a scalar Green's 
function,
\be
\tilde{\bm{\Gamma}}'=(\bm{\nabla\nabla}-\nabla^2\bm{1})G,
\ee
which satisfies
\be
(-\nabla^2+\zeta_m^2+\omega_p^2)G(\rho;\zeta_m)=\delta(\bm{\rho}).
\ee
All of this leads immediately to the following expression for the
internal energy density
\be
u=-\frac{T}{2\pi\rho} \sum_{m=-\infty}^\infty \zeta_m^2e^{i\zeta_m\tau}
e^{-\sqrt{\zeta_m^2+\omega_p^2}\rho},\label{u}
\ee
as one might have anticipated.  (Here, we are inserting a
time-splitting regulator in the same naive manner as before; however, 
see Eq.~(\ref{encutoff}) below---cutoffs in the energy and free energy have
different structures.)

We now expand this to first order in $\omega_p^2$, in view of the remarks at
the beginning of this subsection, and then carry out the Matsubara sum:
\bea
u&=&-\frac{T}{2\pi}\left(\frac1\rho\frac{\partial^2}{\partial\rho^2}+\frac{
\omega_p^2}2\frac\partial{\partial\rho}\right)\left(-2
+\frac1{1-e^{-2\pi T(\rho-i\tau)}}+\frac1{1-e^{-2\pi T(\rho+i\tau)}}\right)
\nonumber\\
&=&\frac1{\pi^2}\frac{3\tau^2-\rho^2}{(\tau^2+\rho^2)^3}+\frac{\pi^2}{15}T^4
-\frac{\omega_p^2}4
\left(\frac1{\pi^2}\frac{\tau^2-\rho^2}{(\tau^2+\rho^2)^2}
+\frac1{3}T^2\right).\label{u0and1}
\eea
For $\omega_p=0$, we  recover twice the previous scalar
 vacuum result (\ref{vacu}).
The resulting entropy density, including the plasma correction, is
($\frac{\partial u}{\partial T}=T\frac{\partial s}{\partial T}$)
\be
s=\frac{4\pi^2}{45}T^3-\frac{\omega_p^2}6 T.\label{s1}
\ee
The correction is indeed very small if $T\gg \omega_p$.

\subsubsection{Weak-coupling, high-temperature, expansion}
\label{sec:pt}
It is easy to carry out this calculation to all orders in
$x=\omega_p/\zeta_1$.  For simplicity, let us suppose $\tau\gg\rho$,  so we
can expand the exponential in Eq.~(\ref{u}):
\be
u=-\frac{T}{2\pi\rho}\sum_{m=-\infty}^\infty \zeta_m^2 e^{i\zeta_m\tau}\left(
1-\rho\sqrt{\zeta_m^2+\omega_p^2}\right).
\ee
The first term is just the derivative of a $\delta$ function,
\be
\sum_{m=-\infty}^\infty \zeta_m^2 e^{i\zeta_m\tau}=-(2\pi)(2\pi T)^2
\delta''(2\pi T\tau),\label{2ndderdelta}
\ee
so may be omitted since we take the {\it limit\/} as $\tau\to 0$.
So with only temporal regulation,
\be
u=\frac{T}{\pi}\sum_{m=1}^\infty \zeta_m^2\sqrt{\zeta_m^2+\omega_p^2}
\cos\zeta_m\tau.
\ee
For $\omega_p\ll \zeta_1$, we expand the square root in a binomial series,
\be
u=\frac{T}\pi\sum_{m=1}^\infty \zeta_m^3\sum_{k=0}^\infty \left(
\begin{array}{c} \frac12\\k \end{array}\right)\left(\frac{\omega_p}{\zeta_m}
\right)^{2k}\cos\zeta_m\tau.
\ee
The first two terms in this series are displayed above in Eq.~(\ref{u0and1}).
The third term is just the expansion of a logarithm,
\be
u^{(2)}=\frac{\omega_p^4}{32\pi^2}\left[\ln\left(1-e^{-i2\pi T\tau}\right)
+\ln\left(1-e^{i2\pi T\tau}\right)\right]=\frac{\omega_p^4}{16\pi^2}
\ln2\pi T\tau. \label{log}
\ee
The remaining terms in the series are finite, so in terms of
$x=\omega_p/(2\pi T)$,
\be
\sum_{k=3}^\infty u^{(k)}=4\pi^{5/2}T^4\sum_{k=3}^\infty \frac{\zeta(2k-3)}{
\Gamma(k+1)\Gamma(3/2-k)}x^{2k}.\label{you}
\ee
In fact, the finite (temperature-dependent) parts of $u$ are given by this
expression for $k=0$ and 1,  while for $k=2$ with the replacement $\zeta(1)\to
-\ln2\pi T\tau$ gives the appropriate $\ln T$ dependence,
seen in Eq.~(\ref{log}). Then using
\be
\frac{\partial u}{\partial T}=T\frac{\partial s}{\partial T},\label{utos}
\ee
we deduce the following expression for the entropy density:
\be
s=4\pi^{5/2}T^3\sum_{k=0}^\infty \frac{(2-k)\zeta(2k-3)}{\Gamma(k+1)\Gamma(
5/2-k)}x^{2k}+s_0,\label{s}
\ee
where $s_0$ is a constant independent of $T$.
Evidently, the radius of convergence of this series is 1.
We can combine Eqs.~(\ref{you}) and (\ref{s}) to give the free energy density,
\be
f=u-Ts=-2\pi^{5/2}T^4\sum_{k=0}^\infty \frac{\zeta(2k-3)}{\Gamma(k+1)\Gamma(
5/2-k)}x^{2k}-Ts_0,
\ee
where again the divergent $\zeta(1)$ term is to be interpreted as a logarithmic
divergence.

\subsubsection{Nonperturbative resummation}
\par Although the above series only converges for $x<1$, it can be analytically continued 
to all positive $x$ by use
of the representation for the Riemann zeta function,
\be
\zeta(s)=\frac1{\Gamma(s)}\int_0^\infty dt\frac{t^{s-1}}{e^t-1},
\ee
which then yields
\be
s=\frac{\omega_p^3}{90\pi x^3}\left\{1-\frac{15}2 x^2-\frac{45}4 x^4
+90 x^3\int_0^\infty \frac{dt}{t}\frac1{e^t-1} J_3(xt)\right\}+s_0,\label{ac}
\ee
where $J_3$ is a Bessel function.
The limit of $s-s_0$ for large $x$ (temperature low compared to the
plasma frequency) is
\be
s-s_0\sim -\frac{\omega_p^3}{6 \pi},\quad x\to\infty,\label{limit}
\ee
so the requirement of the third law of thermodynamics (Nernst's heat theorem)
is that $s_0=\omega_p^3/(6\pi)$.

\begin{figure}
\includegraphics{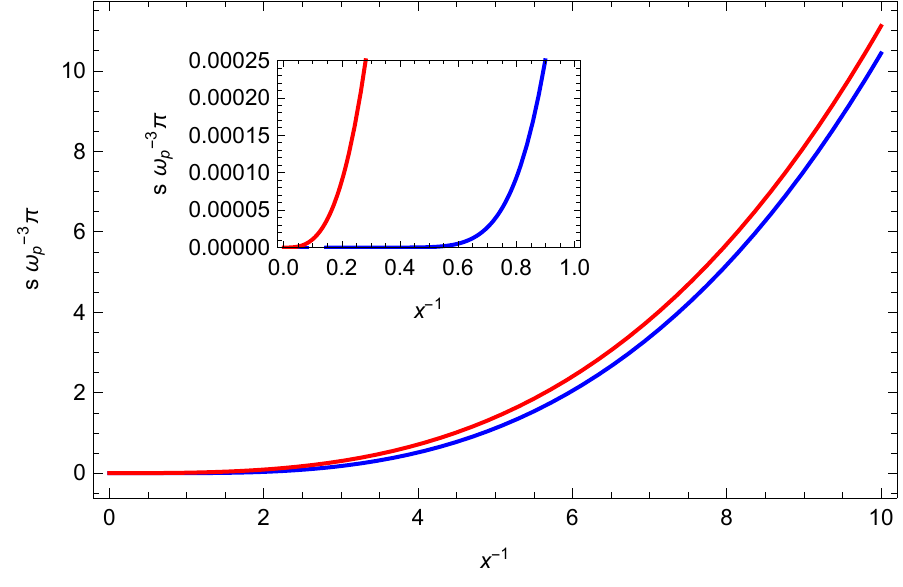}
\caption{\label{compas} Comparison of the vacuum  entropy density (\ref{fands})      
(red curve)with the entropy density including plasma-model dispersion, (\ref{fulls}), 
(blue curve). These are plotted as functions of $x^{-1}=2\pi T/\omega_p$.  Both 
entropy densities tend to zero at zero temperature, and are everywhere positive.  The 
inset shows the behavior for low temperature, and reveals that the plasma-model  entropy 
drops exponentially to zero for relatively large values of the temperature.}
\end{figure}

In fact, if we now make appropriate additions and subtractions to the integrand
in Eq.~(\ref{ac}), we can write the entropy density as
\be
s=\frac{\omega_p^3}\pi\int_0^\infty \frac{dy}y\left(
\frac1{e^{y/x}-1}-\frac{x}{y}+\frac12-\frac{y}{12x}+\frac{y^3}{720 x^3}
\right)J_3(y),\label{fulls}
\ee
where the last term in the integral is defined by analytic continuation.
(For numerical purposes, only the first three subtractions should be
employed, leaving the first term in Eq.~(\ref{ac}).)  Numerically, it appears
that the entropy  is exponentially small in the
$x\to \infty$ limit, that is, there are no power corrections.
Consistent with this,
this limit is actually achieved very early, in the perturbative region,
as Fig.~\ref{compas} shows. This exponential damping is precisely
what is expected in a massive theory---note, here, that the plasma frequency plays the 
role  of a mass.

Formal verification of this can be obtained by constructing a strong-coupling
(low-temperature) expansion for
the quantity in parentheses in Eq.~(\ref{fulls})
by the rest of the Bernoulli expansion,
\be
s=\frac{\omega_p^3}\pi\int_0^\infty\frac{dy}y\sum_{k=3}^\infty \frac{B_{2k}}{2k!}
\left(\frac{y}x\right)^{2k-1} J_3(y).
\ee
We use analytic continuation to define the $y$ integrals:
\be
\int_0^\infty y^{2k-2}J_3(y)=2^{2k-2}\frac{\Gamma(1+k)}{\Gamma(3-k)},
\ee
which vanishes for $k\ge3$.  Thus, the low-temperature expansion of
Eq.~(\ref{fulls}) is zero.

\subsubsection{Strong-coupling, low-temperature,  expansion}
To verify  this limiting behavior, let us try to extract directly the low-temperature 
(large $x$) limit. 
The Euler-Maclaurin formula should be effective in this regard:
\be
\sum_{m=0}^\infty{}' f(m)=\int_0^\infty dm\,f(m)-\sum_{k=1}^\infty
\frac{B_{2k}}{(2k)!}f^{(2k-1)}(0),\label{emsf}
\ee
where the prime means that the $m=0$ term is counted with half weight.
For variety's sake, let's now set $\tau=0$, and keep only the spatial cutoff.
Then the energy density (\ref{u}) is
\be
u=-\frac{T}{\pi\rho}\sum_{m=0}^\infty \zeta_m^2
e^{-\rho\sqrt{\zeta_m^2+\omega_p^2}}
=-\frac{T}{\pi\rho}\left(\frac{\partial^2}{\partial\rho^2}-\omega_p^2\right)
\sum_{m=0}^\infty{}' e^{-\rho\sqrt{\zeta_m^2+\omega_p^2}}.\label{urho}
\ee
The integral term in the Euler-Maclaurin formula gives
\be
\int_0^\infty dm\,e^{-\rho\sqrt{(2\pi m T)^2+\omega_p^2}}=\frac1{2\pi T\rho}
\int_{\omega_p\rho}^\infty du\frac{u}{\sqrt{u^2-\omega_p^2\rho^2}}e^{-u}=
\frac{\omega_p}{2\pi T}K_1(\omega_p\rho).
\ee
Applying the differential operator in Eq.~(\ref{urho}),
\be
\left(\frac{\partial^2}{\partial\rho^2}-\omega_p^2\right)\frac{\omega_p}{
2\pi T}K_1(\omega_p\rho)=\frac{\omega_p^2}{2\pi T\rho}K_2(\omega_p\rho),
\ee
and then using  the small argument expansion of the modified Bessel
function, we obtain for the integral contribution to the internal energy
density
\be
u_{\rm int}=-\frac1{\pi^2}\frac1{\rho^4}+\frac1{4\pi^2}
\frac{\omega_p^2}{\rho^2}+\frac{\omega_p^4}{64\pi^2}\left(-3+4\gamma
+4\ln(\omega_p\rho/2)\right).\label{rhodiv}
\ee
This agrees with the $\tau=0$ divergences displayed in Eq.~(\ref{u0and1}),
while the logarithmic divergence in $\rho$ is the same as that in $\tau$
seen in Eq.~(\ref{log}).

But now it is apparently that $f(m)$ is an even function of $m$, which means
that all the odd derivatives vanish.  Thus, there is no temperature dependence
of the internal energy in the low temperature  limit. (That is, the dependence
 is exponentially small, and nonperturbative in the temperature.)
 This is consistent with the zero value of  the entropy, without power
corrections, found in this
limit above.\footnote{In fact, the Euler-Maclaurin formula is exact if only $n$ terms 
are kept in the Bernoulli sum, and the remainder term is added: 
\be
\frac1{(2n)!}\sum_{k=0}^\infty \int_0^1 dt f^{(2n)}(t+k)B_{2n}(t), 
\ee
where $B_{2n}$ is the Bernoulli polynomial.  Even for $n=1$ it is easily seen
numerically that the contribution to the energy (\ref{urho}) is extremely small if $T$     
is moderately large and $\rho$ is small.}

\subsubsection{Low-temperature asymptotics}
\begin{figure}
    \centering
    \includegraphics{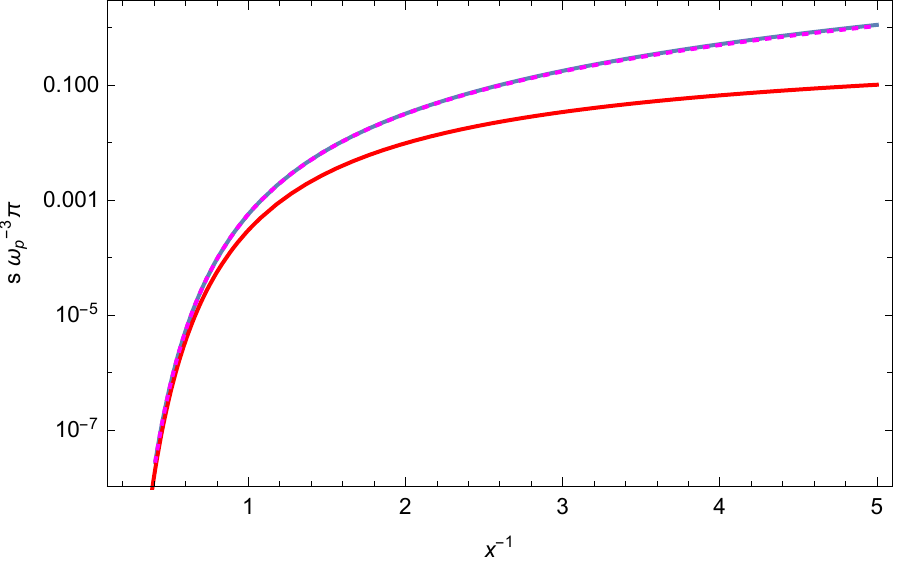}
    \caption{The exact entropy of the blackbody radiation in the plasma model (upper blue 
    curve) compared with the leading low-temperature asymptotic expression (\ref{leadlow})
  (lower red curve) and the numerical integration of the $m=1$ term in Eq.~(\ref{une0}),  
    dotted magenta curve.    Even for $x=0.2$, a rather large value of $T$, the latter is
    less than 5\% low. The cruder approximation (\ref{leadlow}) is only good for rather 
small values of $x^{-1}=2\pi T/\omega_p$.}
    \label{lowfig}
\end{figure}

In fact, we can readily obtain the asymptotic behavior for low temperature.  We rewrite
Eq.~(\ref{urho}) as
\be
u=\frac{T}{2\pi^{3/2}\rho}\left(\frac{\partial^2}{\partial\rho^2}-\omega_p^2\right)
\frac{\partial}{\partial\rho}\rho\int_0^\infty dt\,t^{-3/2}e^{-t}e^{-\rho^2\omega_p^2/(4t)}
\sum_{m=0}^\infty{}'e^{-\rho^2\zeta^2_m/(4t)}.\label{intrep1}
\ee
Now, using the Poisson summation formula, we can recast the $m$ sum into \cite{borwein2}
\be
\frac12\sum_{m=-\infty}^\infty e^{-\rho^2\pi^2 T^2m^2/t}=\frac1{\rho T}\sqrt{\frac{t}
{\pi}}\sum_{m=0}^\infty{}' e^{-m^2 t/(\rho^2 T^2)}.
\ee
Then the $t$ integration gives a Macdonald function:
\be
u=\frac{\omega_p^4}{2\pi^2}\frac1\delta\left(\frac{\partial^2}{\partial\delta^2}-1\right)
\frac\partial{\partial\delta}\sum_{m=-\infty}^\infty K_0(\sqrt{\delta^2+(2\pi m x)^2}),
\ee
where $\delta=\omega_p\rho$.
For $m=0$ this yields exactly the divergent structure (\ref{rhodiv}) as $\delta\to0$.
However, for $m\ne0$, the small $\delta$ limit is finite:
\be
u_{m\ne0}=\frac{\omega_p^4}{\pi^2}\sum_{m=1}^\infty\left[\frac{K_1(z)}{z}
+\frac{3K_2(z)}{z^2}\right],\label{une0}
\quad z=2\pi m x.
\ee
For large $z$, low temperature, this is dominated by the $m=1$ term,
\be
u^{(1)}\sim \frac{\omega_p^4}{4\pi^3}x^{-3/2}e^{-2\pi x},\quad x\to\infty.
\ee  The entropy is obtained by integrating
\be
\frac{\partial s}{\partial x}=\frac{2\pi x}{\omega_p}\frac{\partial u}{\partial x},
\ee
which yields the leading asymptotic approximation
\be
s^{(1)}\sim\frac{\omega_p^3}{2\pi^2}\frac{e^{-2\pi x}}{\sqrt{x}},\quad T\ll\omega_p.
\label{leadlow}
\ee
Comparisons with the exact results obtained from Eq.~(\ref{ac}) are shown in
Fig.~\ref{lowfig}. The exponential suppression of the entropy
for temperatures low compared to the ``mass,'' the plasma frequency, is thus unambiguously
established.

\subsubsection{Free energy}
\label{sec:free}
The above argument is defective in one sense: Because it was based on the internal
energy, it did not determine the constant $s_0$ in the entropy, which had to be
fixed by hand, by requiring that the entropy vanishes at zero temperature, the third
law of thermodynamics.  Therefore, it would appear more satisfactory to start with
the free energy.  However, this is more complex, as we now see.

The free energy is defined by
\be
F=-\frac{T}2\Tr \ln \bm{\Gamma},
\ee
where the trace now includes the sum over Matsubara frequencies.  Now from
Eq.~(\ref{gammainv}), we see that
\be 0=\frac{\partial}{\partial\omega_p^2}\left(\bm{\Gamma}^{-1}\bm{\Gamma}\right)
=-\frac1{\zeta_m^2}
\bm{\Gamma}+\frac{\partial}{\partial \omega_p^2}\ln\bm{\Gamma},
\ee
which says that
\be
\ln\bm{\Gamma}(\omega_p^2)=-\frac1{\zeta_m^2}\int_{\omega_p^2}^\infty
d\omega_p^{\prime2}\,\bm{\Gamma}(\omega_p^{\prime2}). 
\ee
Now the internal energy is proportional to the trace of $\bm{\Gamma}$, so the free
energy can be immediately given after integration on $\omega_p^{\prime2}$ as
\be
f=-\frac{T}{\pi\rho^2}\sum_{m=0}^\infty{}'\left(\sqrt{\zeta_m^2+\omega_p^2}
+\frac1\rho\right)e^{-\rho\sqrt{\zeta_m^2+\omega_p^2}}.\label{fespreg}
\ee
When $\omega_p^2=0$, this directly implies Eq.~(\ref{fands}).

In general, we have to consider $m=0$ separately from the higher Matsubara terms.
(Recall, $m=0$ does not contribute to the internal energy.)  That is
\be
f_{m=0}=\frac{T}{2\pi\rho^2}\left(\frac\partial{\partial\rho}-\frac1\rho\right)e^{-\rho
\omega_p}\sim -\frac{T}{2\pi\rho^3}+\frac{T\omega_p^2}{4\pi\rho}
-\frac{T\omega_p^3}{6\pi},\quad \rho\to0.\label{mequals0}
\ee
The $T$-dependent divergent terms must be cancelled by the remainder of the Matsubara
series.  As before, we can proceed perturbatively in powers of $\omega_p$. The zeroth
order term is
\be
f^{(0)}_{m\ne0} = \frac{T}{\pi\rho^2}\left(\frac{\partial}{\partial\rho}-\frac1\rho\right)
\sum_{m=1}^\infty e^{-2\pi m T\rho}\sim -\frac1{\pi^2\rho^2}+\frac{T}{2\pi\rho^3}-
\frac{\pi^2 T^4}{45},\quad\rho\to0
\ee
where the first and last terms reproduce Eq.~(\ref{fands}) and the middle term cancels
the $\omega_p$-independent term in Eq.~(\ref{mequals0}).

The term of order $\omega_p^2$ is
\be
f^{(1)}_{m\ne0}=\frac{\omega_p^2 T}{2\pi\rho}\sum_{m=1}^\infty e^{-\rho\zeta_m}\sim
\frac{\omega_p^2}{4\pi \rho^2}-\frac{\omega_p^2 T}{4\pi \rho}+\frac{\omega_p^2 T^2}{12},
\quad\rho\to 0.
\ee
The second term here cancels the second term in Eq.~(\ref{mequals0}), and we are left with
the structure seen in Eqs.~(\ref{u0and1}) and (\ref{s1}).  As for the term of
order $\omega_p^4$, we have
\be
f_{m\ne0}^{(2)}=-\frac{\omega_p^4}{16\pi^2}\sum_{m=1}^\infty \frac1m e^{-\delta m/x}\sim
\frac{\omega_p^4}{16\pi^2}\ln 2\pi T\rho,
\ee
again as anticipated.

Since this approach seems a bit complicated, we will reinsert a temporal
regulator, which simplifies the calculation. But we have already achieved our goal:
It is clear that the perturbative expansion for $m\ne0$ contributions involves only even 
powers of $\omega_p$, so the only term of order $\omega_p^3$ is that remaining in 
Eq.~(\ref{mequals0}). This term, linear in the temperature, corresponds to the constant 
term in the entropy, undetermined by the previous analysis:
\be
s_0=\frac{\omega_p^3}{6\pi}.
\ee

Naively inserting the temporal regulator into Eq.~(\ref{fespreg}), we write
\be
f=\frac{T}{2\pi\rho^2}\left(\frac\partial{\partial\rho}-\frac1\rho\right)
\sum_{m=-\infty}^\infty e^{-\rho\sqrt{\zeta_m^2+\omega_p^2}} e^{i\zeta_m\tau}.
\ee
Now with $\tau\ne0$, we can expand in $\rho$, and after omitting terms involving
$\delta$ functions in $\tau$, find
\be
f=-\frac{T}{6\pi}\sum_{m=-\infty}^\infty (\zeta_m^2+\omega_p^2)^{3/2} e^{i\zeta_m\tau}.
\ee
Now, this is readily expanded in powers of $x=\omega_p/\zeta_1$.  Following the by-now
familiar procedure, we find the first three divergent terms:
\begin{subequations}
\bea
f^{(0)}&=&-\frac1{\pi^2\tau^4}-\frac{\pi^2T^4}{45},\label{f0}\\
f^{(1)}&=&\frac{\omega_p^2}{4\pi^2\tau^2}+\frac{\omega_p^2}{12}T^2,\label{f1}\\
f^{(2)}&=&\frac{\omega_p^4}{16\pi^2}\ln 2\pi T\tau.\label{f2}
\eea
\end{subequations}
The remaining terms are finite. The only odd term in $\omega_p$ comes from the $m=0$ term:
\be
f_{m=0}=-\frac{\omega_p^3 T}{6\pi},
\ee
which corresponds precisely to the constant term in the entropy.
Then, the higher terms in $\omega_p$ are
\begin{subequations}
\be
f^{(k\ge3)}=\frac{8\pi^2T^4}{3}\frac{\Gamma(5/2)\zeta(2k-3)}{\Gamma(k+1)\Gamma(5/2-k)}
x^{2k},
\ee
corresponding to the entropy term
\be
s^{(k\ge3)}=4\pi^{5/2}T^3\frac{(2-k)\zeta(2k-3)}{\Gamma(k+1)\Gamma(5/2-k)}x^{2k}.
\ee
\end{subequations}
The temperature-dependence of $f$, and the entropy are precisely those found previously in
Sec.~\ref{sec:pt}.  The cutoff terms, however, are off by $-\frac13$ and $-1$ for the
quartically divergent and quadratically terms, respectively.
This is due, as explained in
Ref.~\cite{parashar2017electromagnetic}, to the fact that an exponential temporal
cutoff in the
free energy corresponds to a more elaborate form for the internal energy:
\be
e^{i\zeta\tau}\to\frac{e^{i\zeta\tau}-1}{i\zeta\tau}\label{encutoff}
\ee
This precisely accounts for the discrepant factors in the zero-temperature divergences.

\section{Conclusions}
\label{C}

Previously \cite{Milton2017CasimirSelf,milton2019remarks,Li2021Negativity},
we analyzed the self-entropy of a macroscopic sphere.  But viewed
from far away, a compact object appears to be a particle.  So,
in this paper, we examine the question of self-entropy from the 
nanoparticle perspective.
By nanoparticle, we mean that the  size of the particle
is small compared to any other length scale, such as the inverse temperature.
This self-entropy exhibits surprising features, especially its negativity
for weak coupling to the electromagnetic field.  We approach this question
by first extracting the polarizabilities of a nanoparticle through consideration of
its effect on the
scattering of the electromagnetic field.  From these polarizabilities, expressed in terms
of the reflection coefficients, we can compute the free energy and entropy by summing
over Matsubara frequencies.  The results apply to low temperatures, compared to the inverse
size of the nanoparticle.  (For a 100 nm particle, the temperature for which $aT=1$ is 
24,000 K.)

\par Specifically, we illustrate these ideas by investigating two models. First, for a 
$\delta$-function spherical shell in which the permittivity on the surface is transverse 
and 
represented by the plasma model, we obtain results for the self-entropy which give the 
leading low-temperature response, linear in the coupling, for the TE and TM modes; these 
self-entropies are precisely those found
earlier \cite{Milton2017CasimirSelf}.  The second model, a dielectric/diamagnetic ball 
without dispersion, involves more subtleties. The scattering-derived polarizabilities give
 a contribution to the free energy which is linear in the susceptibility, in the dilute 
limit, which therefore violates the expected connection between Casimir and van der Waals 
forces \cite{milton2020self}.  

This discrepancy is resolved by the subtraction of the ``bulk contribution,'' which is 
the contribution to the Casimir free energy corresponding to
the nonscattering Green's functions due to the medium either inside or
outside the spherical boundary filling the whole space. (Such a contribution only removes 
the vacuum contribution for a hollow spherical shell.) Performing this subtraction, we 
remove the linear term, and recover the negative self-entropy found two decades ago
\cite{barton2001perturbative,bartoniv,nesterenko2001casimir}.
That self-entropy is reproduced again in Appendices \ref{appa}--\ref{appc}.

However, the reader might well object to the fact that the known
energies at zero temperature are not reproduced by the procedure proposed here.
After all, in second order in the susceptibility, the finite part
of the energy of the dielectric ball
is as given by the first term in Eq.~(\ref{eqPE.DB.3}).
But here the
only finite terms in the free energy are those going like a power of the temperature,
yielding a finite self-entropy.  Furthermore, the divergent term found
here for the dielectric ball is the difference between the $\tau$-dependent
terms in Eqs.~(\ref{eqPE.DB.2}) 
and (\ref{eqBFE.8}),  
\be
F_{\rm div}=\frac{7a^3}{6\pi \tau^4}(\varepsilon-1)^2,
\ee
while the energy found some forty years ago \cite{milton1980} is less singular,
\be
E_{\rm div}=\frac{a^2}{8\tau^3}(\varepsilon-1)^2.
\ee
One might think that the reason for not seeing the 
temperature-independent finite terms is that extraction of these requires
more sophisticated analytic continuation techniques, such as zeta-function
regularization, which sweep divergences under the rug.\footnote{Both the zero-temperature
free energy of the dilute dielectric ball, and its first temperature correction, are
    rederived in Appendix 
    \ref{appa}. But again, the magic of analytic 
    continuation plays an essential role.}
Apparently, the point-particle viewpoint
is only effective in extracting the temperature-dependent part of the free energy, and
therefore the entropy, but not the finite or divergent parts.  
As argued
in Appendix \ref{Appb}, it would be anticipated that the point-particle viewpoint
being explored here is only effective in extracting extensive quantities, proportional
to the volume of the particle, and not contributions to the free energy going like lower
powers of the radius of the particle.

On the other hand, we  now  understand the appearance
 of  negative self-entropies found for objects, be they macroscropic
dielectric balls or small nanoparticles.
The approach presented here reveals the negative entropy as arising from an interaction 
between the nanoparticle and the blackbody radiation.  
As discussed in Appendix~\ref{Appb}, the free energy of the blackbody
radiation in vacuum, and in the material of the dielectric ball, both without
interaction (that is, the ``nonscattering'' part), must be subtracted from the
total free energy to obtain the self-free energy, or the interaction
free energy. This precisely corresponds to our bulk subtraction.
The resulting interaction entropy can be, in fact, negative.
(It bears a resemblance to, but in general,
is not the same as, the negative of the relative entropy discussed in  
Ref.~\cite{breuer}.)
The total entropy, of course, has the bulk, blackbody entropies included, so is always 
positive.

\begin{acknowledgments}
The work of KAM was supported by a grant from the US National Science Foundation, grant
number 2008417.  It is a pleasure to acknowledge the collaborative assistance of
 Stephen Fulling and Iver Brevik.  This paper reflects solely the authors' personal
 opinions and does not represent the opinions of the authors' employers, present and
 past, in any way.
\end{acknowledgments}

\appendix

\section{Bulk Subtraction at Finite Temperature}
\label{appa}
It might be thought that the bulk subtraction, which we used to derive the entropy of the 
dielectric/diamagnetic ball, need only be employed at zero temperature. After all, the bulk
 divergences occur only at zero temperature. But this is not correct, because the reason 
for the subtraction is not primarily to remove divergences, but to remove the effects of a 
homogeneous medium, the non-scattering contribution
of the Green's function to the internal energy.  We see this in the general formula 
(\ref{generalu}), from which the zero-temperature form for the energy is 
obtained, for example, by the Euler-Maclaurin summation formula.  This involves all
$m$, so the replacement of the Green's dyadic by its scattering part must be applied
universally.  The subtraction is not required in order to make the energy finite (which it 
is not, even after this truncation of the Green's function), 
but because it is necessary physically:
the Casimir-Lifshitz energy must arise, at least in the dilute approximation, 
from the summation of van der Waals or Casimir-Polder interactions,
which are quadratic in the polarizability, so the Casimir-Lifshitz free energy at
arbitrary temperature must
start from order $(\varepsilon-1)^2$ for a dilute dielectric sphere.
All of this was recognized by Dzyaloshinskii,  Lifshitz, and  Pitaevskii \cite{dlp},
and constitutes the ``Lifshitz subtraction.''

\subsection{Casimir-Polder Free Energy of Two Dilute Dielectric Slabs}
We can demonstrate this explicitly by considering the classic Lifshitz configuration 
\cite{lifshitz}
of two parallel dielectric slabs separated by a vacuum region of width $a$.  In the
limit where the two dielectric media are dilute, we can compute this in two ways:
either by summing the van der Waals interactions between the various atoms constituting
the slabs, or by taking the dilute limit of the Lifshitz free energy.  Let's consider
the high-temperature limit, and suppose that the Casimir-Polder interaction 
\cite{casimirpolder} between 
atoms with polarizability $\alpha_1$ and $\alpha_2$ separated by a distance $r$ is
in the classical limit
\be
V=-C T \frac{\alpha_1\alpha_2}{r^6}, \label{vdwhiT}
\ee a structure required dimensionally, where $C$ is a dimensionless
constant to be determined.
The free energy of the two slabs is obtained by summing the interactions between the
two slabs, which are supposed to be homogeneous, with number densities $N_1$ and $N_2$.
The free energy is
\be
F_{\rm CP}=- C T N_1\alpha_1 N_2\alpha_2\int_{-\infty}^0 dz\int_a^\infty dz'
\int(d\mathbf{r}_\perp)(d\mathbf{r'}_\perp)
\left[(\mathbf{r_\perp-r'_\perp})^2+(z-z')^2\right]^{-3}.
\ee
The integrations are elementary, leading to the free energy per unit area
\be
\frac{F_{\rm CP}}A=-C T\frac\pi{12}\frac{N_1N_1\alpha_1\alpha_2}{a^2}.\label{cpt}
\ee

The bulk-subtracted Lifshitz pressure between the two slabs is \cite{lifshitz,sdm1978}
\be
p=-\frac{T}{8\pi}\sum_{m=-\infty}^\infty\int_0^\infty dk^2 \,2\kappa_3
\left(\frac1{\Delta^E}+\frac1{\Delta^H}\right),
\ee
for the configuration where the top and bottom slabs are denoted 1 and 2, and the
intermediate vacuum region is denoted 3.  The denominators are, in terms of the inverses
of the reflection coefficients at the interfaces,
\begin{subequations}
\bea
\Delta^E&=&\frac{\kappa_3+\kappa_1}{\kappa_3-\kappa_1}\frac{\kappa_3+\kappa_2}{\kappa_3-
\kappa_2}e^{2\kappa_3 a}-1,\\
\Delta^H&=&\frac{\kappa_3/\varepsilon_3+\kappa_1/\varepsilon_1}
{\kappa_3/\varepsilon_3-\kappa_1/\varepsilon_1}\frac{\kappa_3/\varepsilon_3
+\kappa_2/\varepsilon_2}{\kappa_3/\varepsilon_3-
\kappa_2/\varepsilon_2}e^{2\kappa_3 a}-1,
\eea
\end{subequations}
where 
\be \kappa_3=\kappa=\sqrt{k^2+\zeta_m^2},\quad
\kappa_{1,2}=\sqrt{k^2+\varepsilon_{1,2}(\zeta_m)\zeta_m^2}.
\ee
When $\varepsilon_{1,2}-1$ are both small, this is readily expanded to the leading
order:  introducing the variable $u=\kappa/\zeta_m$, 
we have
\be
p=-\frac{T}{16\pi}\sum_{m=0}^\infty\!{}'\zeta_m^3(\varepsilon_1(\zeta_m)-1)
(\varepsilon_2(\zeta_m)-1)\int_1^\infty \frac{du}{u^2}[1+(1-2u^2)^2]e^{-2\zeta_m u a}.
\ee
Now we are interested in high temperature, which corresponds to $m=0$, so we take the 
limit
of small $\zeta_m$.  In that limit, the $u$ integral is asymptotically 
$(\zeta_m a)^{-3}$,
and then the $m=0$ term in the free energy, obtained by integrating 
\be p=-\frac{\partial}{\partial a}\frac{F}A
\ee
is, in terms of the static dielectric constants
\be
\frac{F}A=-\frac{T}{64\pi}(\varepsilon_1-1)(\varepsilon_2-1)\frac1{a^2},
\quad aT\gg1.\label{lfe}
\ee

These two expressions, Eq.~(\ref{lfe}) and Eq.~(\ref{cpt}), should be identical.
They are, when we recall the connection between the dielectric constant and the
polarizability:
\be
\varepsilon=1+N \alpha,
\ee
and this determines the constant in the van der Waals potential to be $C=3/(4\pi)^2$.

\subsection{Casimir-Polder Free Energy of Dilute Dielectric Ball}
We can straightforwardly extend this argument to sum the Casimir-Polder energies
of the atomic constituents of a dilute dielectric sphere.  This should correspond
to the free energy of such an object. Such a calculation, in fact,  is the same idea
as that exploited by Barton \cite{bartoniv}, although the calculational details are
different, as are the results.

The free energy of interaction between the constituents of isotropic polarizability 
$\alpha$ is given by \cite{milton2015negative}
\be
F_{CP}=-\frac12 \frac{N^2\alpha^2}{(4\pi)^2}
\int(d\mathbf{r})(d\mathbf{r'}) \frac1{4\pi \rho^7} f(y),
\ee
where $y=4\pi \rho T$, $\rho=\sqrt{r^2-r^{\prime 2}-\mathbf{r\cdot r'}}$, and
\begin{subequations}
\be
f(y)=yD_T \coth\frac{y}2,
\ee with
\be D_T=3-3T\partial_T+\frac54 T^2\partial_T^2-\frac14 T^3\partial_T^3+\frac1{16}
T^4\partial_T^4.
\ee
\end{subequations}
(Note, at high $T$, where $\coth y/2$ tends to 1, this yields Eq.~(\ref{vdwhiT}) 
with $C=3/(4\pi)^2$.)
Here $N$ is the number density of polarizable constituents.  The integrals extend over
the volume of the sphere wherein the constituents reside.  Since $\rho$ never gets large,
to extract the low-temperature behavior, we can expand the hyperbolic cotangent 
for small argument:
\be
\coth z =\frac1z+\frac{z}3-\frac{z^3}{45}+\frac{2z^5}{45}+\dots.
\ee
Consider the first term in the expansion.  Since the differential operator $D_T$
applied to
$1/T$ yields $\frac{23}{2T}$, we are left with evaluating the radial integral
\be
\int(d\mathbf{r})(d\mathbf{r'}) \frac1{\rho^7}.
\ee
This integral is divergent, because the coordinates can be coincident, but a method of
extracting the finite part was described in Refs.~\cite{milton1998vdw,miltonbook}. Let's consider
the generic integral
\be
I(\gamma)=\int(d\mathbf{r})(d\mathbf{r'})\frac1{\rho^\gamma}.
\ee
The result is, for $\gamma<3$,
\be
I(\gamma)=\frac{128 \pi^{2} 
2^{-\gamma}}{(6-\gamma)(4-\gamma)(3-\gamma)}a^{6-\gamma}.\label{Igamma}
\ee
This integral is well defined if $\gamma$ is 7, so by analytic
continuation we immediately obtain the zero temperature free energy
\be
E_0=\frac{23}{1536}\frac{(\varepsilon-1)^2}{\pi a},
\ee
where we identify $\varepsilon-1= N \alpha$.  This is the well-known result
\cite{milton1998vdw},  seen in Eq.~(\ref{eqPE.DB.3}),

As for the low temperature corrections, it is apparent that the differential operator
$D_T$
annihilates a term linear in $T$, so the $O(z)$ term in the expansion of the cotangent
is without effect because $I(5)$ is finite.  
$D_T$ has the same effect on the $T^3$ term.  But that corresponds to 
$\gamma=3$, where the volume integral (\ref{Igamma}) has a pole, so we must take a
limit. Write $y^3\to y^{3-s}$ where $s\to 0$.  
So then
\be
D_T T^{3-s}\to-\frac78 s T^3,\quad s\to 0,
\ee
while 
\be
I(3+s)\to -\frac{16 \pi^2}{3s} a^3,
\ee
so the product is finite as $s\to 0$, and supplying the other factors, we obtain
for the temperature correction to the free energy
\be
\Delta F_T= \frac{7}{270}(\varepsilon-1)^2 (\pi a)^3 T^4,
\ee
precisely the result (\ref{dbfe})
given by Nesterenko et al. \cite{nesterenko2001casimir}
but without the additional $T^3$ term found by Barton \cite{bartoniv}.

\section{Sign of the Interaction Entropy for a Dielectric Ball}
\label{Appb}

In this appendix, we explore the sign of the interaction entropy for a dielectric ball, 
using the Clausius-Mossotti relation \cite{schwinger2000classical}.
We begin with a critical appraisal of the approximation of the finite-size nanoparticle 
by a point particle with the same polarizability, and an appreciation of the limitations 
of this approximation.

The use of Eq.~(\ref{eqPE.2}) in Eq.~(\ref{eqPE.1}) implies a resummation of the usual
perturbative expansion to all orders in $\varepsilon -1$, to extract all contributions to
leading order in the particle size, which, in total, give rise to its polarizability. By
construction, therefore, this procedure can only produce the {\it extensive\/} 
contribution to the free energy of the nanoparticle, that is, that part of the free 
energy that is proportional to the volume of the nanoparticle. It cannot capture any 
other contributions to the free energy that are not extensive, such as the surface term 
obtained by Barton \cite{barton2001perturbative}, or the zero-temperature term displayed 
in Eq.~(\ref{dbfe}), which is proportional to the inverse radius of the nanoparticle 
\cite{nesterenko2001casimir,barton2001perturbative,bartoniv,milton1998vdw,%
brevik1999identity}, or, indeed, any contributions that are of higher order in the 
polarizability. Likewise, only the extensive contribution to the corresponding derived 
entropy can result. This is a fundamental limitation of the approximation 
employed.\footnote{In Section II, we considered only the
single-scattering approximation. It may be possible to capture information about the 
{\it shape\/} of the nanoparticle, rather than just its size, by extending the approach 
to include successively higher orders of scattering. This may reveal the dependence on 
its surface area, its mean extrinsic curvature, {\it etc.}.}

In this approximation, the free energy, $F$, of the nanoparticle may be regarded as a 
linear function of the extensive variable, $V$, the volume of the nanoparticle, and a 
possibly nonlinear function of the intensive variable, $N$, the number density of its 
polarizable constituents. Below, we explore the assembly of the nanoparticle from its 
initially widely-dispersed constituents, and so are interested in how the free energy 
changes with the volume, while keeping the number of constituents fixed. Since
\begin{equation}
\frac{dF}{dV}=\frac{\partial F}{\partial V} - \frac{N}{V} \frac{\partial F}{\partial N}=
\frac{F}{V} - \frac{N}{V} \frac{\partial F}{\partial N},
\label{B1}
\end{equation}
it follows that $\frac{dF}{dV}=0$, that is, the free energy of the nanoparticle is 
invariant under this change in volume, if and only if it is also a linear function of 
$N$. In fact, this is the case for the Clausius-Mossotti relation, which expresses the 
polarizability of a dielectric ball as a linear function of the number
density, and polarizability, of its constituents. It should be noted that linearity in 
the number density of the constituents does not preclude interaction between these 
constituents. Indeed, the Clausius-Mossotti relation derives from the inclusion of the 
effect on the induced dipole moment of each constituent of the induced electric field due
to the induced dipole moments of the other constituents. This interaction between the
constituents is reflected in the nonlinear dependence of the 
polarizability on $\varepsilon -1$.

To be specific, let us therefore assume that the permeability $\mu=1$ and that the system 
consists of a dielectric ball, $B$, of radius $a$ and permittivity $\varepsilon$, which is 
composed of polarizable constituents of number density $N$ and polarizability $\alpha_c$, 
and which is surrounded by vacuum. Let us also consider the corresponding reference system
 which consists of the dilated ball, $B_{\lambda}$, of radius $\lambda a$ and permittivity
 $\varepsilon_{\lambda}$, wherein the polarizable constituents have number density 
$N_{\lambda}\equiv \frac{N}{\lambda^3}$, where $\lambda >1$, which is also surrounded by 
vacuum. From the Clausius-Mossotti relation, we immediately obtain the relation between 
the corresponding permittivities:
\begin{equation}
\frac{\varepsilon-1}{\varepsilon+2}=\frac{N\alpha_c}{3}=\frac{\lambda^3 N_{\lambda}
\alpha_c}{3}=\lambda^3 \left(\frac{\varepsilon_{\lambda}-1}{\varepsilon_{\lambda}+2}
\right).
\label{B2}
\end{equation}

The total free energy of the original system may be constructed from three components,
\begin{equation}
F_{\text{tot}}=F_{\text{vac}}^{{\mathbb R}^3}+\left(F_{\text{med}}^B-F_{\text{vac}}^B
\right)+F_{\text{int}}^B,
\label{B3}
\end{equation}
using an obvious notation: $F_{\text{med}}^B-F_{\text{vac}}^B$ is the change in the bulk 
free energy of the ball resulting from replacement of vacuum by medium in its interior, 
which arises from the corresponding change in the non-scattering part of the Green's 
function, and $F_{\text{int}}^B$ is the interaction free energy between the interior of 
the ball (medium) and its exterior (vacuum), which arises from the scattering part of the 
corresponding Green's function. Likewise, the total free energy of the reference system may
 be written as 
\begin{equation}
F_{\text{tot}}^{\lambda}=F_{\text{vac}}^{{\mathbb R}^3}+\left(F_{\text{med}}^{B_{\lambda}}-
F_{\text{vac}}^{B_{\lambda}}\right)+F_{\text{int}}^{B_{\lambda}}.\label{B4}
\end{equation}

The change in the bulk free energy of the dilated ball is
\begin{equation}
F_{\text{med}}^{B_{\lambda}}-F_{\text{vac}}^{B_{\lambda}}=\int_{B_{\lambda}} 
d\mathbf{r}\left(\varepsilon_{\lambda}^{\frac32}-1\right) f_0(\tau, T)=\frac{4}{3}\pi a^3
\lambda^3 \left[\left(\frac{1+\frac{2N\alpha_c}{3\lambda^3}}{1
-\frac{N\alpha_c}{3\lambda^3}}\right)^{\!\!\frac32}-1\right] f_0(\tau, T), 
\label{B5}
\end{equation}
which becomes, in the limit $\lambda\to \infty$, 
\begin{equation}
\lim_{\lambda\to\infty} 
\left(F_{\text{med}}^{B_{\lambda}}-F_{\text{vac}}^{B_{\lambda}}\right)=\frac43 \pi a^3 
\,\frac92 \left(\frac{\varepsilon -1}{\varepsilon+2}\right) f_0(\tau, T),
\label{B6}
\end{equation}
where 
\begin{equation}
f_0(\tau, T)\equiv -\frac{1}{\pi^2\tau^4}-\frac{\pi^2T^4}{45}
\label{B7}
\end{equation}
is the vacuum free energy density (\ref{f0})
under the temporal regulator $\tau$. As expected, this 
is simply the sum over the finite number, $\frac43 \pi a^3 N$, of the then infinitely 
separated and therefore non-interacting polarizable constituents, each of which 
contributes $\frac32 \alpha_c f_0(\tau, T)$ to the free energy. Correspondingly, in this 
limit, the interaction free energy of the dilated ball must vanish, as it then fills all 
space and has no exterior, so $\lim_{\lambda\to\infty} F_{\text{int}}^{B_{\lambda}}=0$. 
(See below.)

Thus,
\begin{equation}
\left(F_{\text{med}}^B-F_{\text{vac}}^B\right)
-\lim_{\lambda\to\infty} 
\left(F_{\text{med}}^{B_{\lambda}}-F_{\text{vac}}^{B_{\lambda}}\right)
=\frac{4}{3}\pi a^3 \left[\varepsilon^{\frac32}-1-\frac92
\left(\frac{\varepsilon-1}{\varepsilon+2}\right)\right] f_0(\tau, T).
\label{B8}
\end{equation}
This expression represents the change in the total bulk free energy when the ball is 
assembled from its initially maximally dispersed constituents, and its sign is determined 
from that of 
\begin{equation}
g(\varepsilon)\equiv \varepsilon^{\frac32}-1-\frac92
\left(\frac{\varepsilon-1}{\varepsilon+2}\right).\label{B9}\end{equation}
It is easily verified that $g(1)=0$ and $g(\varepsilon) > 0$ for $\varepsilon  > 0$,
$\varepsilon \ne 1$. Therefore, the change in the total bulk free energy is negative,
and, correspondingly, from Eq.~(\ref{B7}), the change in the total bulk entropy is
positive.

However, Eq.~(\ref{B8}) may also be written as
$\left(F_{\text{tot}}-F_{\text{int}}^B\right) - 
\left(F_{\text{tot}}^{\infty}-\lim_{\lambda \to \infty} 
F_{\text{int}}^{B_{\lambda}}\right)=-F_{\text{int}}^B$, since $\lim_{\lambda \to 
\infty}F_{\text{int}}^{B_{\lambda}}=0$ and $F_{\text{tot}}=F_{\text{tot}}^{\infty}$, the 
latter because of the linear dependence of the free energy of the dielectric ball on the 
number density of its polarizable constituents, inherited from the Clausius-Mossotti 
relation, as described in Eq.~(\ref{B1}). Thus, 
\begin{subequations}
\begin{equation}
F_{\text{int}}^B=\frac43 \pi a^3 \left[\frac92 \left(\frac{\varepsilon -1}{\varepsilon +2}\right)-\left(\varepsilon^{\frac32}-1\right)\right] f_0(\tau, T) =-\frac43 \pi a^3 g(\varepsilon) \,f_0(\tau, T) > 0\label{B15}
\end{equation}
and, correspondingly,
\begin{equation}
S_{\text{int}}^B= \left[\frac92 \left(\frac{\varepsilon -1}{\varepsilon +2}\right)-\left(\varepsilon^{\frac32}-1\right)\right] \frac{2 t^3}{135} =-g(\varepsilon) \frac{2t^3}{135} < 0, \quad t=2\pi a T.
\label{B17}
\end{equation}
\end{subequations}
That is, the ball has a positive interaction free energy and a
corresponding negative interaction entropy.

From Eq.~(\ref{B15}), we can demonstrate self-consistency by noting
that the interaction
free energy of the dilated ball vanishes as the dilation factor goes to 
infinity:
\begin{equation}
    F_{\rm int}^{B_\lambda}=-\frac{4\pi(\lambda a)^3}{3}\left[\varepsilon_\lambda^{3/2}-1
    -\frac92\frac{\varepsilon_\lambda-1}{\varepsilon_\lambda+2}\right]f_0(\tau,T)\to
    -\frac{\pi a^3}2\frac{N^2\alpha_c^2}{\lambda^3}f_0(\tau,T)\to 0,\quad \lambda\to\infty.
\end{equation}

The important point to note here is that, because, under the Clausius-Mossotti relation,
the total free energy of the system is unchanged when the ball is assembled from its
initially maximally dispersed and therefore non-interacting constituents, the 
{\it decrease\/} in the total {\it bulk\/} free energy resulting from the assembly is 
exactly 
offset by a corresponding {\it increase\/} in the initially vanishing {\it interaction\/}
free energy, resulting here in a positive interaction free energy and a
corresponding negative interaction entropy for the ball. The assembly of the ball from
its constituents would otherwise be a spontaneous process, decreasing the total bulk free
energy and increasing the total bulk entropy. However, the assembly itself creates the
differentiation between the interior (medium) and the exterior (vacuum) of the ball, and 
the resulting macroscopic inhomogeneity in the permittivity gives rise to the scattering 
part of the Green's function and the corresponding inhibiting positive interaction free 
energy and negative interaction entropy. In this point of view, the appearance of a 
negative interaction entropy is seen as a natural and  inevitable consequence of the 
creation of the scattering contribution to the Green's function through the assembly 
process. (Of course, this conclusion only holds for $\mu=1$.)

In spite of the negative interaction entropy, insertion of the particle into the vacuum 
always decreases (if $\varepsilon>1$) the total free energy and increases
 the total entropy of the system:
\begin{subequations}
\begin{equation}
F_{\text{tot}}-F_{\text{vac}}^{{\mathbb R}^3}=\left(F_{\text{med}}^B-F_{\text{vac}}^B
\right) + F_{\text{int}}^B=6 \pi a^3  \left(\frac{\varepsilon -1}{\varepsilon +2}\right) 
f_0(\tau, T) <0 \label{B12}
\end{equation}
and
\begin{equation}
S_{\text{tot}}-S_{\text{vac}}^{{\mathbb R}^3}=\left(S_{\text{med}}^B-S_{\text{vac}}^B
\right) + S_{\text{int}}^B=\left(\frac{\varepsilon -1}{\varepsilon +2}\right) 
\frac{t^3}{15} > 0.
\label{B13}
\end{equation}
\end{subequations}

\begin{figure}
    \centering
    \includegraphics{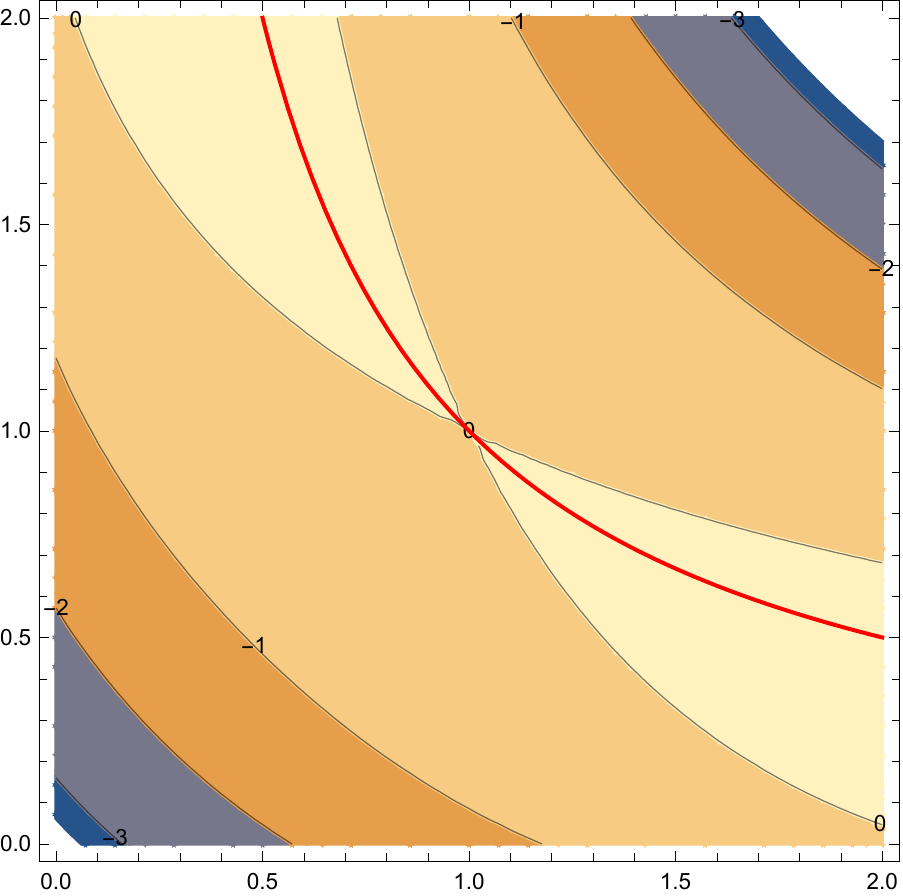}
    \caption{The entropy given in terms of 
 $-g(\varepsilon,\mu)$ in Eq.~(\ref{sepsmu}) for 
    $\varepsilon$ and $\mu$ between $0$ and $2$.  The contour lines denote the
boundaries of regions where  the function has values between the designated integers.  
The positive entropy region is colored
lightest  (yellow), roughly following the $\varepsilon\mu=1$ hyperbola shown
in red.}  
    \label{posent}
\end{figure}

More generally, the interaction entropy could be of either sign.
If the ball has permeability as well as permittivity, the interaction
entropy becomes
\begin{equation}
S_{\rm int}=-g(\varepsilon,\mu)\frac{2t^3}{135},
\quad g(\varepsilon,\mu)=(\varepsilon\mu)^{3/2}-1
-\frac92\left(\frac{\varepsilon-1}{\varepsilon+2}
+\frac{\mu-1}{\mu+2}\right).\label{sepsmu}
\end{equation}
A perfectly conducting spherical shell corresponds to $\varepsilon\to\infty$,
$\mu\to 0$ so that $\varepsilon\mu\to1$.  (The latter corresponds to the interior
of the shell being vacuum as in the exterior.)  In this way the familiar result
(\ref{pcs}) emerges,
\begin{equation}
S_{\rm PCS}=\frac{t^3}{30},\label{pcs2}
\end{equation}
which is positive.  But it is not necessary to go to the limit
to see the change in sign. Although the interaction
entropy (\ref{sepsmu}) is nonpositive
whenever $\varepsilon$ and $\mu$ are both $\ge1$, if one of these is less than unity,
there is a region of positive entropy, as illustrated in Fig.~\ref{posent}.

\section{Direct Calculation of Low-Temperature Correction to Interaction Free Energy}
\label{appc}
It is easy to obtain the general results for the low-temperature correction to the
free energy, and the low-temperature entropy, directly from the zero-temperature
expressions derived many years ago \cite{milton1980,milton1997casimir,miltonbook}.
The pressure on a dielectric/diamagnetic ball is given there by
\begin{equation}
p=\frac1{2a^3}\int_{-\infty}^\infty \frac{d\zeta}{2\pi}e^{i\zeta\tau}\sum_{l=1}^\infty
\frac{2l+1}{4\pi} \left[f_l(x)-f^{(0)}_l(x)\right],
\end{equation}
where the contribution from the bulk subtraction is
\begin{subequations}
\begin{equation}
    f^{(0)}_l(x)=2 x[s'_l(x)e_l'(x)-e_l(x)s_l''(x)]-2 x'[s'_l(x')e_l'(x')-e_l(x')s_l''(x')],
\end{equation}
where $x=|\zeta|a$, $x'=\sqrt{\varepsilon\mu}x$,
and the scattering contribution is
\begin{equation}
  f_l(x)=  x\frac{d}{dx}D_l(x),\quad D_l(x)=[s_l(x')e_l'(x)-s_l'(x')e_l(x)]^2-\xi^2
    [s_l(x')e_l'(x)+s_l'(x')e_l(x)]^2,
\end{equation}
\end{subequations}
with 
\begin{equation}
\xi=\frac{\sqrt{\frac{\varepsilon}{\mu}}-1}{\sqrt{\frac{\varepsilon}{\mu}}+1}.
\end{equation}

To extend these old results to finite temperature, we simply replace the integral
over Euclidean frequencies by a sum over Matsubara frequencies:
\be
\int_{-\infty}^\infty \frac{d\zeta}{2\pi}\to T\sum_{m=-\infty}^\infty,\quad \zeta\to
\zeta_m=2\pi m T.
\ee
For low temperatures, we can evaluate the sum by using the Euler-Maclaurin sum formula
(\ref{emsf}).  The integral there, of course, corresponds to the original
zero-temperature result, which contains all the divergences.  The sum over Bernoulli
numbers contains the finite temperature corrections, which must be understood as
an asymptotic series.  So we need the odd derivatives of $f_l$ and $f_l^{(0)}$ at zero
to find the temperature corrections, which requires computing the odd terms
in the power series expansion of these functions about the origin.  
The leading power, as would
have been anticipated \cite{Milton2017CasimirSelf}, occurs only for $l=1$:
\begin{equation}
f_1(x)\sim -2x^3\left(\frac{\varepsilon-1}{\varepsilon+2}+\frac{\mu-1}{\mu+2}\right),
\quad
f_1^{(0)}(x)\sim\frac49x^3\left(1-(\varepsilon\mu)^{3/2}\right),\quad x\ll1.
\end{equation}
Supplying the remaining factors, we immediately obtain the small temperature correction
to the pressure,
\begin{equation}
\Delta P=-\frac{\pi^2T^4}{45}g(\varepsilon,\mu).
\end{equation}
Since $4\pi a^2 p=-\frac{\partial}{\partial a}F$, the temperature correction to
the free energy is just as found above,
\begin{equation}
\Delta F=\frac{4\pi a^3}{3}\frac{\pi^2 T^4}{45}g(\varepsilon,\mu),
\end{equation}
and the entropy (\ref{sepsmu}) follows.  This elementary calculation should have been done
40 years ago.

\bibliography{ref}

\begin{thebibliography}{54}%
\makeatletter
\providecommand \@ifxundefined [1]{%
 \@ifx{#1\undefined}
}%
\providecommand \@ifnum [1]{%
 \ifnum #1\expandafter \@firstoftwo
 \else \expandafter \@secondoftwo
 \fi
}%
\providecommand \@ifx [1]{%
 \ifx #1\expandafter \@firstoftwo
 \else \expandafter \@secondoftwo
 \fi
}%
\providecommand \natexlab [1]{#1}%
\providecommand \enquote  [1]{``#1''}%
\providecommand \bibnamefont  [1]{#1}%
\providecommand \bibfnamefont [1]{#1}%
\providecommand \citenamefont [1]{#1}%
\providecommand \href@noop [0]{\@secondoftwo}%
\providecommand \href [0]{\begingroup \@sanitize@url \@href}%
\providecommand \@href[1]{\@@startlink{#1}\@@href}%
\providecommand \@@href[1]{\endgroup#1\@@endlink}%
\providecommand \@sanitize@url [0]{\catcode `\\12\catcode `\$12\catcode
  `\&12\catcode `\#12\catcode `\^12\catcode `\_12\catcode `\%12\relax}%
\providecommand \@@startlink[1]{}%
\providecommand \@@endlink[0]{}%
\providecommand \url  [0]{\begingroup\@sanitize@url \@url }%
\providecommand \@url [1]{\endgroup\@href {#1}{\urlprefix }}%
\providecommand \urlprefix  [0]{URL }%
\providecommand \Eprint [0]{\href }%
\providecommand \doibase [0]{http://dx.doi.org/}%
\providecommand \selectlanguage [0]{\@gobble}%
\providecommand \bibinfo  [0]{\@secondoftwo}%
\providecommand \bibfield  [0]{\@secondoftwo}%
\providecommand \translation [1]{[#1]}%
\providecommand \BibitemOpen [0]{}%
\providecommand \bibitemStop [0]{}%
\providecommand \bibitemNoStop [0]{.\EOS\space}%
\providecommand \EOS [0]{\spacefactor3000\relax}%
\providecommand \BibitemShut  [1]{\csname bibitem#1\endcsname}%
\let\auto@bib@innerbib\@empty
\bibitem [{\citenamefont {Bezerra}\ \emph {et~al.}(2004)\citenamefont
  {Bezerra}, \citenamefont {Klimchitskaya}, \citenamefont {Mostepanenko},\ and\
  \citenamefont {Romero}}]{bezerra2004violation}%
  \BibitemOpen
  \bibfield  {author} {\bibinfo {author} {\bibfnamefont {V.~B.}\ \bibnamefont
  {Bezerra}}, \bibinfo {author} {\bibfnamefont {G.~L.}\ \bibnamefont
  {Klimchitskaya}}, \bibinfo {author} {\bibfnamefont {V.~M.}\ \bibnamefont
  {Mostepanenko}}, \ and\ \bibinfo {author} {\bibfnamefont {C.}~\bibnamefont
  {Romero}},\ }\bibfield  {title} {\enquote {\bibinfo {title} {{Violation of
  the Nernst heat theorem in the theory of the thermal Casimir force between
  Drude metals}},}\ }\href@noop {} {\bibfield  {journal} {\bibinfo  {journal}
  {Phys. Rev. A}\ }\textbf {\bibinfo {volume} {69}},\ \bibinfo {pages} {022119}
  (\bibinfo {year} {2004})}\BibitemShut {NoStop}%
\bibitem [{\citenamefont {Klimchitskaya}\ and\ \citenamefont
  {Korikov}(2015)}]{klimchitskaya2015casimir}%
  \BibitemOpen
  \bibfield  {author} {\bibinfo {author} {\bibfnamefont {G.~L.}\ \bibnamefont
  {Klimchitskaya}}\ and\ \bibinfo {author} {\bibfnamefont {C.~C.}\ \bibnamefont
  {Korikov}},\ }\bibfield  {title} {\enquote {\bibinfo {title} {Casimir entropy
  for magnetodielectrics},}\ }\href@noop {} {\bibfield  {journal} {\bibinfo
  {journal} {J. Phys. Condens. Matter}\ }\textbf {\bibinfo {volume} {27}},\
  \bibinfo {pages} {214007} (\bibinfo {year} {2015})}\BibitemShut {NoStop}%
\bibitem [{\citenamefont {Korikov}(2016)}]{korikov2016casimir}%
  \BibitemOpen
  \bibfield  {author} {\bibinfo {author} {\bibfnamefont {C.~C.}\ \bibnamefont
  {Korikov}},\ }\bibfield  {title} {\enquote {\bibinfo {title} {Casimir entropy
  for ferromagnetic materials},}\ }\href@noop {} {\bibfield  {journal}
  {\bibinfo  {journal} {Int. J. Mod. Phys. A}\ }\textbf {\bibinfo {volume}
  {31}},\ \bibinfo {pages} {1641036} (\bibinfo {year} {2016})}\BibitemShut
  {NoStop}%
\bibitem [{\citenamefont {Klimchitskaya}\ and\ \citenamefont
  {Mostepanenko}(2017)}]{klimchitskaya2017low}%
  \BibitemOpen
  \bibfield  {author} {\bibinfo {author} {\bibfnamefont {G.~L.}\ \bibnamefont
  {Klimchitskaya}}\ and\ \bibinfo {author} {\bibfnamefont {V.~M.}\ \bibnamefont
  {Mostepanenko}},\ }\bibfield  {title} {\enquote {\bibinfo {title}
  {{Low-temperature behavior of the Casimir free energy and entropy of metallic
  films}},}\ }\href@noop {} {\bibfield  {journal} {\bibinfo  {journal} {Phys.
  Rev. A}\ }\textbf {\bibinfo {volume} {95}},\ \bibinfo {pages} {012130}
  (\bibinfo {year} {2017})}\BibitemShut {NoStop}%
\bibitem [{\citenamefont {Brevik}\ \emph {et~al.}(2006)\citenamefont {Brevik},
  \citenamefont {Ellingsen},\ and\ \citenamefont {Milton}}]{brevik2006thermal}%
  \BibitemOpen
  \bibfield  {author} {\bibinfo {author} {\bibfnamefont {I.}~\bibnamefont
  {Brevik}}, \bibinfo {author} {\bibfnamefont {S.~A.}\ \bibnamefont
  {Ellingsen}}, \ and\ \bibinfo {author} {\bibfnamefont {K.~A.}\ \bibnamefont
  {Milton}},\ }\bibfield  {title} {\enquote {\bibinfo {title} {{Thermal
  corrections to the Casimir effect}},}\ }\href@noop {} {\bibfield  {journal}
  {\bibinfo  {journal} {New J. Phys.}\ }\textbf {\bibinfo {volume} {8}},\
  \bibinfo {pages} {236} (\bibinfo {year} {2006})}\BibitemShut {NoStop}%
\bibitem [{\citenamefont {Milton}\ \emph {et~al.}(2012)\citenamefont {Milton},
  \citenamefont {Brevik},\ and\ \citenamefont {Ellingsen}}]{milton2012thermal}%
  \BibitemOpen
  \bibfield  {author} {\bibinfo {author} {\bibfnamefont {K.~A.}\ \bibnamefont
  {Milton}}, \bibinfo {author} {\bibfnamefont {I.}~\bibnamefont {Brevik}}, \
  and\ \bibinfo {author} {\bibfnamefont {S.~A.}\ \bibnamefont {Ellingsen}},\
  }\bibfield  {title} {\enquote {\bibinfo {title} {{Thermal issues in Casimir
  forces between conductors and semiconductors}},}\ }\href@noop {} {\bibfield
  {journal} {\bibinfo  {journal} {Phys. Scr.}\ }\textbf {\bibinfo {volume}
  {2012}},\ \bibinfo {pages} {014070} (\bibinfo {year} {2012})}\BibitemShut
  {NoStop}%
\bibitem [{\citenamefont {Decca}\ \emph {et~al.}(2005)\citenamefont {Decca},
  \citenamefont {L\'{o}pez}, \citenamefont {Fischbach}, \citenamefont
  {Klimchitskaya}, \citenamefont {Krause},\ and\ \citenamefont
  {Mostepanenko}}]{Decca2005Precise}%
  \BibitemOpen
  \bibfield  {author} {\bibinfo {author} {\bibfnamefont {R.~S.}\ \bibnamefont
  {Decca}}, \bibinfo {author} {\bibfnamefont {D.}~\bibnamefont {L\'{o}pez}},
  \bibinfo {author} {\bibfnamefont {E.}~\bibnamefont {Fischbach}}, \bibinfo
  {author} {\bibfnamefont {G.~L.}\ \bibnamefont {Klimchitskaya}}, \bibinfo
  {author} {\bibfnamefont {D.~E.}\ \bibnamefont {Krause}}, \ and\ \bibinfo
  {author} {\bibfnamefont {V.~M.}\ \bibnamefont {Mostepanenko}},\ }\bibfield
  {title} {\enquote {\bibinfo {title} {{Precise comparison of theory and new
  experiment for the Casimir force leads to stronger constraints on thermal
  quantum effects and long-range interactions}},}\ }\href@noop {} {\bibfield
  {journal} {\bibinfo  {journal} {Ann. Phys. (N. Y.)}\ }\textbf {\bibinfo
  {volume} {318}},\ \bibinfo {pages} {37--80} (\bibinfo {year}
  {2005})}\BibitemShut {NoStop}%
\bibitem [{\citenamefont {Banishev}\ \emph {et~al.}(2013)\citenamefont
  {Banishev}, \citenamefont {Klimchitskaya}, \citenamefont {Mostepanenko},\
  and\ \citenamefont {Mohideen}}]{Banishev2013Casimir}%
  \BibitemOpen
  \bibfield  {author} {\bibinfo {author} {\bibfnamefont {A.~A.}\ \bibnamefont
  {Banishev}}, \bibinfo {author} {\bibfnamefont {G.~L.}\ \bibnamefont
  {Klimchitskaya}}, \bibinfo {author} {\bibfnamefont {V.~M.}\ \bibnamefont
  {Mostepanenko}}, \ and\ \bibinfo {author} {\bibfnamefont {U.}~\bibnamefont
  {Mohideen}},\ }\bibfield  {title} {\enquote {\bibinfo {title} {{Casimir
  interaction between two magnetic metals in comparison with nonmagnetic test
  bodies}},}\ }\href@noop {} {\bibfield  {journal} {\bibinfo  {journal} {Phys.
  Rev. B}\ }\textbf {\bibinfo {volume} {88}},\ \bibinfo {pages} {5514--5518}
  (\bibinfo {year} {2013})}\BibitemShut {NoStop}%
\bibitem [{\citenamefont {Bimonte}\ \emph {et~al.}(2016)\citenamefont
  {Bimonte}, \citenamefont {Lopez},\ and\ \citenamefont
  {Decca}}]{Bimonte2016Isoelectronic}%
  \BibitemOpen
  \bibfield  {author} {\bibinfo {author} {\bibfnamefont {G.}~\bibnamefont
  {Bimonte}}, \bibinfo {author} {\bibfnamefont {D.}~\bibnamefont {Lopez}}, \
  and\ \bibinfo {author} {\bibfnamefont {R.~S.}\ \bibnamefont {Decca}},\
  }\bibfield  {title} {\enquote {\bibinfo {title} {{Isoelectronic determination
  of the thermal Casimir force}},}\ }\href@noop {} {\bibfield  {journal}
  {\bibinfo  {journal} {Phys. Rev. B}\ }\textbf {\bibinfo {volume} {93}},\
  \bibinfo {pages} {184434} (\bibinfo {year} {2016})}\BibitemShut {NoStop}%
\bibitem [{\citenamefont {Liu}\ \emph {et~al.}(2019)\citenamefont {Liu},
  \citenamefont {Xu}, \citenamefont {Klimchitskaya}, \citenamefont
  {Mostepanenko},\ and\ \citenamefont {Mohideen}}]{Liu2019Examining}%
  \BibitemOpen
  \bibfield  {author} {\bibinfo {author} {\bibfnamefont {M.}~\bibnamefont
  {Liu}}, \bibinfo {author} {\bibfnamefont {J.}~\bibnamefont {Xu}}, \bibinfo
  {author} {\bibfnamefont {G.~L.}\ \bibnamefont {Klimchitskaya}}, \bibinfo
  {author} {\bibfnamefont {V.~M.}\ \bibnamefont {Mostepanenko}}, \ and\
  \bibinfo {author} {\bibfnamefont {U.}~\bibnamefont {Mohideen}},\ }\bibfield
  {title} {\enquote {\bibinfo {title} {{Examining the Casimir puzzle with
  upgraded technique and advanced surface cleaning}},}\ }\href@noop {}
  {\bibfield  {journal} {\bibinfo  {journal} {Phys. Rev. B}\ }\textbf {\bibinfo
  {volume} {100}},\ \bibinfo {pages} {081406} (\bibinfo {year}
  {2019})}\BibitemShut {NoStop}%
\bibitem [{\citenamefont {Klimchitskaya}\ and\ \citenamefont
  {Mostepanenko}(2021)}]{klimchitskaya2021}%
  \BibitemOpen
  \bibfield  {author} {\bibinfo {author} {\bibfnamefont {G.~L.}\ \bibnamefont
  {Klimchitskaya}}\ and\ \bibinfo {author} {\bibfnamefont {V.~M.}\ \bibnamefont
  {Mostepanenko}},\ }\bibfield  {title} {\enquote {\bibinfo {title} {Casimir
  effect for magnetic media: Spatially nonlocal response to the off-shell
  quantum fluctuations},}\ }\href@noop {} {\bibfield  {journal} {\bibinfo
  {journal} {Phys. Rev. D}\ }\textbf {\bibinfo {volume} {104}},\ \bibinfo
  {pages} {085001} (\bibinfo {year} {2021})}\BibitemShut {NoStop}%
\bibitem [{\citenamefont {Sushkov}\ \emph {et~al.}(2011)\citenamefont
  {Sushkov}, \citenamefont {Kim}, \citenamefont {Dalvit},\ and\ \citenamefont
  {Lamoreaux}}]{Sushkov2011Observation}%
  \BibitemOpen
  \bibfield  {author} {\bibinfo {author} {\bibfnamefont {A.~O.}\ \bibnamefont
  {Sushkov}}, \bibinfo {author} {\bibfnamefont {W.~J.}\ \bibnamefont {Kim}},
  \bibinfo {author} {\bibfnamefont {D.~A.~R}\ \bibnamefont {Dalvit}}, \ and\
  \bibinfo {author} {\bibfnamefont {S.~K.}\ \bibnamefont {Lamoreaux}},\
  }\bibfield  {title} {\enquote {\bibinfo {title} {{Observation of the thermal
  Casimir force}},}\ }\href@noop {} {\bibfield  {journal} {\bibinfo  {journal}
  {Nat. Phys.}\ }\textbf {\bibinfo {volume} {7}},\ \bibinfo {pages} {230--233}
  (\bibinfo {year} {2011})}\BibitemShut {NoStop}%
\bibitem [{\citenamefont {Garcia-Sanchez}\ \emph {et~al.}(2012)\citenamefont
  {Garcia-Sanchez}, \citenamefont {Fong}, \citenamefont {Bhaskaran},
  \citenamefont {Lamoreaux},\ and\ \citenamefont {Hong}}]{Garcia2012Casimir}%
  \BibitemOpen
  \bibfield  {author} {\bibinfo {author} {\bibfnamefont {D.}~\bibnamefont
  {Garcia-Sanchez}}, \bibinfo {author} {\bibfnamefont {K.~Y.}\ \bibnamefont
  {Fong}}, \bibinfo {author} {\bibfnamefont {H.}~\bibnamefont {Bhaskaran}},
  \bibinfo {author} {\bibfnamefont {S.}~\bibnamefont {Lamoreaux}}, \ and\
  \bibinfo {author} {\bibfnamefont {X.~T.}\ \bibnamefont {Hong}},\ }\bibfield
  {title} {\enquote {\bibinfo {title} {Casimir force and in situ surface
  potential measurements on nanomembranes.}}\ }\href@noop {} {\bibfield
  {journal} {\bibinfo  {journal} {Phys. Rev. Lett.}\ }\textbf {\bibinfo
  {volume} {109}},\ \bibinfo {pages} {027202} (\bibinfo {year}
  {2012})}\BibitemShut {NoStop}%
\bibitem [{\citenamefont {Bezerra}\ \emph
  {et~al.}(2002{\natexlab{a}})\citenamefont {Bezerra}, \citenamefont
  {Klimchitskaya},\ and\ \citenamefont
  {Mostepanenko}}]{bezerra2002thermodynamical}%
  \BibitemOpen
  \bibfield  {author} {\bibinfo {author} {\bibfnamefont {V.~B.}\ \bibnamefont
  {Bezerra}}, \bibinfo {author} {\bibfnamefont {G.~L.}\ \bibnamefont
  {Klimchitskaya}}, \ and\ \bibinfo {author} {\bibfnamefont {V.~M.}\
  \bibnamefont {Mostepanenko}},\ }\bibfield  {title} {\enquote {\bibinfo
  {title} {{Thermodynamical aspects of the Casimir force between real metals at
  nonzero temperature}},}\ }\href@noop {} {\bibfield  {journal} {\bibinfo
  {journal} {Phys. Rev. A}\ }\textbf {\bibinfo {volume} {65}},\ \bibinfo
  {pages} {052113} (\bibinfo {year} {2002}{\natexlab{a}})}\BibitemShut
  {NoStop}%
\bibitem [{\citenamefont {Bezerra}\ \emph
  {et~al.}(2002{\natexlab{b}})\citenamefont {Bezerra}, \citenamefont
  {Klimchitskaya},\ and\ \citenamefont
  {Mostepanenko}}]{bezerra2002correlation}%
  \BibitemOpen
  \bibfield  {author} {\bibinfo {author} {\bibfnamefont {V.~B.}\ \bibnamefont
  {Bezerra}}, \bibinfo {author} {\bibfnamefont {G.~L.}\ \bibnamefont
  {Klimchitskaya}}, \ and\ \bibinfo {author} {\bibfnamefont {V.~M.}\
  \bibnamefont {Mostepanenko}},\ }\bibfield  {title} {\enquote {\bibinfo
  {title} {Correlation of energy and free energy for the thermal {Casimir}
  force between real metals},}\ }\href@noop {} {\bibfield  {journal} {\bibinfo
  {journal} {Phys. Rev. A}\ }\textbf {\bibinfo {volume} {66}},\ \bibinfo
  {pages} {062112} (\bibinfo {year} {2002}{\natexlab{b}})}\BibitemShut
  {NoStop}%
\bibitem [{\citenamefont {Canaguier-Durand}\ \emph {et~al.}(2010)\citenamefont
  {Canaguier-Durand}, \citenamefont {Neto}, \citenamefont {Lambrecht},\ and\
  \citenamefont {Reynaud}}]{canaguier2010thermal}%
  \BibitemOpen
  \bibfield  {author} {\bibinfo {author} {\bibfnamefont {A.}~\bibnamefont
  {Canaguier-Durand}}, \bibinfo {author} {\bibfnamefont {P.~A.~M.}\
  \bibnamefont {Neto}}, \bibinfo {author} {\bibfnamefont {A.}~\bibnamefont
  {Lambrecht}}, \ and\ \bibinfo {author} {\bibfnamefont {S.}~\bibnamefont
  {Reynaud}},\ }\bibfield  {title} {\enquote {\bibinfo {title} {{Thermal
  Casimir effect for Drude metals in the plane-sphere geometry}},}\ }\href@noop
  {} {\bibfield  {journal} {\bibinfo  {journal} {Phys. Rev. A}\ }\textbf
  {\bibinfo {volume} {82}},\ \bibinfo {pages} {012511} (\bibinfo {year}
  {2010})}\BibitemShut {NoStop}%
\bibitem [{\citenamefont {Rodriguez-Lopez}(2011)}]{rodriguez2011casimir}%
  \BibitemOpen
  \bibfield  {author} {\bibinfo {author} {\bibfnamefont {P.}~\bibnamefont
  {Rodriguez-Lopez}},\ }\bibfield  {title} {\enquote {\bibinfo {title} {Casimir
  energy and entropy in the sphere-sphere geometry},}\ }\href@noop {}
  {\bibfield  {journal} {\bibinfo  {journal} {Phys. Rev. B}\ }\textbf {\bibinfo
  {volume} {84}},\ \bibinfo {pages} {075431} (\bibinfo {year}
  {2011})}\BibitemShut {NoStop}%
\bibitem [{\citenamefont {Ingold}\ \emph {et~al.}(2015)\citenamefont {Ingold},
  \citenamefont {Umrath}, \citenamefont {Hartmann}, \citenamefont
  {Gu{\'e}rout}, \citenamefont {Lambrecht}, \citenamefont {Reynaud},\ and\
  \citenamefont {Milton}}]{ingold2015geometric}%
  \BibitemOpen
  \bibfield  {author} {\bibinfo {author} {\bibfnamefont {G.}~\bibnamefont
  {Ingold}}, \bibinfo {author} {\bibfnamefont {S.}~\bibnamefont {Umrath}},
  \bibinfo {author} {\bibfnamefont {M.}~\bibnamefont {Hartmann}}, \bibinfo
  {author} {\bibfnamefont {R.}~\bibnamefont {Gu{\'e}rout}}, \bibinfo {author}
  {\bibfnamefont {A.}~\bibnamefont {Lambrecht}}, \bibinfo {author}
  {\bibfnamefont {S.}~\bibnamefont {Reynaud}}, \ and\ \bibinfo {author}
  {\bibfnamefont {K.~A.}\ \bibnamefont {Milton}},\ }\bibfield  {title}
  {\enquote {\bibinfo {title} {Geometric origin of negative {Casimir}
  entropies: A scattering-channel analysis},}\ }\href@noop {} {\bibfield
  {journal} {\bibinfo  {journal} {Phys. Rev. E}\ }\textbf {\bibinfo {volume}
  {91}},\ \bibinfo {pages} {033203} (\bibinfo {year} {2015})}\BibitemShut
  {NoStop}%
\bibitem [{\citenamefont {Milton}\ \emph {et~al.}(2015)\citenamefont {Milton},
  \citenamefont {Gu{\'e}rout}, \citenamefont {Ingold}, \citenamefont
  {Lambrecht},\ and\ \citenamefont {Reynaud}}]{milton2015negative}%
  \BibitemOpen
  \bibfield  {author} {\bibinfo {author} {\bibfnamefont {K.~A.}\ \bibnamefont
  {Milton}}, \bibinfo {author} {\bibfnamefont {R.}~\bibnamefont {Gu{\'e}rout}},
  \bibinfo {author} {\bibfnamefont {G.}~\bibnamefont {Ingold}}, \bibinfo
  {author} {\bibfnamefont {A.}~\bibnamefont {Lambrecht}}, \ and\ \bibinfo
  {author} {\bibfnamefont {S.}~\bibnamefont {Reynaud}},\ }\bibfield  {title}
  {\enquote {\bibinfo {title} {{Negative Casimir entropies in nanoparticle
  interactions}},}\ }\href@noop {} {\bibfield  {journal} {\bibinfo  {journal}
  {J. Phys.: Condens. Matter}\ }\textbf {\bibinfo {volume} {27}},\ \bibinfo
  {pages} {214003} (\bibinfo {year} {2015})}\BibitemShut {NoStop}%
\bibitem [{\citenamefont {Umrath}\ \emph {et~al.}(2015)\citenamefont {Umrath},
  \citenamefont {Hartmann}, \citenamefont {Ingold},\ and\ \citenamefont
  {Neto}}]{Disentangling2015Umrath}%
  \BibitemOpen
  \bibfield  {author} {\bibinfo {author} {\bibfnamefont {S.}~\bibnamefont
  {Umrath}}, \bibinfo {author} {\bibfnamefont {M.}~\bibnamefont {Hartmann}},
  \bibinfo {author} {\bibfnamefont {G.}~\bibnamefont {Ingold}}, \ and\ \bibinfo
  {author} {\bibfnamefont {P.~A.~M.}\ \bibnamefont {Neto}},\ }\bibfield
  {title} {\enquote {\bibinfo {title} {{Disentangling geometric and dissipative
  origins of negative Casimir entropies}},}\ }\href@noop {} {\bibfield
  {journal} {\bibinfo  {journal} {Phys. Rev. E}\ }\textbf {\bibinfo {volume}
  {92}},\ \bibinfo {pages} {042125} (\bibinfo {year} {2015})}\BibitemShut
  {NoStop}%
\bibitem [{\citenamefont {Milton}\ \emph
  {et~al.}(2017{\natexlab{a}})\citenamefont {Milton}, \citenamefont {Li},
  \citenamefont {Kalauni}, \citenamefont {Parashar}, \citenamefont
  {Gu{\'e}rout}, \citenamefont {Ingold}, \citenamefont {Lambrecht},\ and\
  \citenamefont {Reynaud}}]{milton2017negative}%
  \BibitemOpen
  \bibfield  {author} {\bibinfo {author} {\bibfnamefont {K.~A.}\ \bibnamefont
  {Milton}}, \bibinfo {author} {\bibfnamefont {Y.}~\bibnamefont {Li}}, \bibinfo
  {author} {\bibfnamefont {P.}~\bibnamefont {Kalauni}}, \bibinfo {author}
  {\bibfnamefont {P.}~\bibnamefont {Parashar}}, \bibinfo {author}
  {\bibfnamefont {R.}~\bibnamefont {Gu{\'e}rout}}, \bibinfo {author}
  {\bibfnamefont {G.}~\bibnamefont {Ingold}}, \bibinfo {author} {\bibfnamefont
  {A.}~\bibnamefont {Lambrecht}}, \ and\ \bibinfo {author} {\bibfnamefont
  {S.}~\bibnamefont {Reynaud}},\ }\bibfield  {title} {\enquote {\bibinfo
  {title} {Negative entropies in {Casimir} and {Casimir}-{Polder}
  interactions},}\ }\href@noop {} {\bibfield  {journal} {\bibinfo  {journal}
  {Forts. Phys.}\ }\textbf {\bibinfo {volume} {65}},\ \bibinfo {pages}
  {1600047} (\bibinfo {year} {2017}{\natexlab{a}})}\BibitemShut {NoStop}%
\bibitem [{\citenamefont {Li}\ \emph {et~al.}(2016)\citenamefont {Li},
  \citenamefont {Milton}, \citenamefont {Kalauni},\ and\ \citenamefont
  {Parashar}}]{Li2016Casimir}%
  \BibitemOpen
  \bibfield  {author} {\bibinfo {author} {\bibfnamefont {Y.}~\bibnamefont
  {Li}}, \bibinfo {author} {\bibfnamefont {K.~A.}\ \bibnamefont {Milton}},
  \bibinfo {author} {\bibfnamefont {P.}~\bibnamefont {Kalauni}}, \ and\
  \bibinfo {author} {\bibfnamefont {P.}~\bibnamefont {Parashar}},\ }\bibfield
  {title} {\enquote {\bibinfo {title} {Casimir self-entropy of an
  electromagnetic thin sheet},}\ }\href@noop {} {\bibfield  {journal} {\bibinfo
   {journal} {Phys. Rev. D}\ }\textbf {\bibinfo {volume} {94}},\ \bibinfo
  {pages} {085010} (\bibinfo {year} {2016})}\BibitemShut {NoStop}%
\bibitem [{\citenamefont {Milton}\ \emph
  {et~al.}(2017{\natexlab{b}})\citenamefont {Milton}, \citenamefont {Kalauni},
  \citenamefont {Parashar},\ and\ \citenamefont {Li}}]{Milton2017CasimirSelf}%
  \BibitemOpen
  \bibfield  {author} {\bibinfo {author} {\bibfnamefont {K.~A.}\ \bibnamefont
  {Milton}}, \bibinfo {author} {\bibfnamefont {P.}~\bibnamefont {Kalauni}},
  \bibinfo {author} {\bibfnamefont {P.}~\bibnamefont {Parashar}}, \ and\
  \bibinfo {author} {\bibfnamefont {Y.}~\bibnamefont {Li}},\ }\bibfield
  {title} {\enquote {\bibinfo {title} {Casimir self-entropy of a spherical
  electromagnetic $\ensuremath{\delta}$-function shell},}\ }\href@noop {}
  {\bibfield  {journal} {\bibinfo  {journal} {Phys. Rev. D}\ }\textbf {\bibinfo
  {volume} {96}},\ \bibinfo {pages} {085007} (\bibinfo {year}
  {2017}{\natexlab{b}})}\BibitemShut {NoStop}%
\bibitem [{\citenamefont {Bordag}(2018)}]{bordag2018free}%
  \BibitemOpen
  \bibfield  {author} {\bibinfo {author} {\bibfnamefont {M.}~\bibnamefont
  {Bordag}},\ }\bibfield  {title} {\enquote {\bibinfo {title} {Free energy and
  entropy for thin sheets},}\ }\href@noop {} {\bibfield  {journal} {\bibinfo
  {journal} {Phys. Rev. D}\ }\textbf {\bibinfo {volume} {98}},\ \bibinfo
  {pages} {085010} (\bibinfo {year} {2018})}\BibitemShut {NoStop}%
\bibitem [{\citenamefont {Bordag}\ and\ \citenamefont
  {Kirsten}(2018)}]{bordag2018entropy}%
  \BibitemOpen
  \bibfield  {author} {\bibinfo {author} {\bibfnamefont {M.}~\bibnamefont
  {Bordag}}\ and\ \bibinfo {author} {\bibfnamefont {K.}~\bibnamefont
  {Kirsten}},\ }\bibfield  {title} {\enquote {\bibinfo {title} {On the entropy
  of a spherical plasma shell},}\ }\href@noop {} {\bibfield  {journal}
  {\bibinfo  {journal} {J. Phys. A}\ }\textbf {\bibinfo {volume} {51}},\
  \bibinfo {pages} {455001} (\bibinfo {year} {2018})}\BibitemShut {NoStop}%
\bibitem [{\citenamefont {Milton}\ \emph {et~al.}(2019)\citenamefont {Milton},
  \citenamefont {Kalauni}, \citenamefont {Parashar},\ and\ \citenamefont
  {Li}}]{milton2019remarks}%
  \BibitemOpen
  \bibfield  {author} {\bibinfo {author} {\bibfnamefont {K.~A.}\ \bibnamefont
  {Milton}}, \bibinfo {author} {\bibfnamefont {P.}~\bibnamefont {Kalauni}},
  \bibinfo {author} {\bibfnamefont {P.}~\bibnamefont {Parashar}}, \ and\
  \bibinfo {author} {\bibfnamefont {Y.}~\bibnamefont {Li}},\ }\bibfield
  {title} {\enquote {\bibinfo {title} {Remarks on the {Casimir} self-entropy of
  a spherical electromagnetic $\delta$-function shell},}\ }\href@noop {}
  {\bibfield  {journal} {\bibinfo  {journal} {Phys. Rev. D}\ }\textbf {\bibinfo
  {volume} {99}},\ \bibinfo {pages} {045013} (\bibinfo {year}
  {2019})}\BibitemShut {NoStop}%
\bibitem [{\citenamefont {Li}\ \emph {et~al.}(2021)\citenamefont {Li},
  \citenamefont {Milton}, \citenamefont {Parashar},\ and\ \citenamefont
  {Hong}}]{Li2021Negativity}%
  \BibitemOpen
  \bibfield  {author} {\bibinfo {author} {\bibfnamefont {Y.}~\bibnamefont
  {Li}}, \bibinfo {author} {\bibfnamefont {K.~A.}\ \bibnamefont {Milton}},
  \bibinfo {author} {\bibfnamefont {P.}~\bibnamefont {Parashar}}, \ and\
  \bibinfo {author} {\bibfnamefont {L.~J.}\ \bibnamefont {Hong}},\ }\bibfield
  {title} {\enquote {\bibinfo {title} {Negativity of the {Casimir} self-entropy
  in spherical geometries},}\ }\href@noop {} {\bibfield  {journal} {\bibinfo
  {journal} {Entropy}\ }\textbf {\bibinfo {volume} {23}},\ \bibinfo {pages}
  {214} (\bibinfo {year} {2021})}\BibitemShut {NoStop}%
\bibitem [{\citenamefont {Milton}\ \emph {et~al.}(2020)\citenamefont {Milton},
  \citenamefont {Parashar}, \citenamefont {Brevik},\ and\ \citenamefont
  {Kennedy}}]{milton2020self}%
  \BibitemOpen
  \bibfield  {author} {\bibinfo {author} {\bibfnamefont {K.~A.}\ \bibnamefont
  {Milton}}, \bibinfo {author} {\bibfnamefont {P.}~\bibnamefont {Parashar}},
  \bibinfo {author} {\bibfnamefont {I.}~\bibnamefont {Brevik}}, \ and\ \bibinfo
  {author} {\bibfnamefont {G.}~\bibnamefont {Kennedy}},\ }\bibfield  {title}
  {\enquote {\bibinfo {title} {{Self-stress on a dielectric ball and
  Casimir--Polder forces}},}\ }\href@noop {} {\bibfield  {journal} {\bibinfo
  {journal} {Ann. Phys. (N. Y.)}\ }\textbf {\bibinfo {volume} {412}},\ \bibinfo
  {pages} {168008} (\bibinfo {year} {2020})}\BibitemShut {NoStop}%
\bibitem [{\citenamefont {Milton}\ and\ \citenamefont
  {Brevik}(2018)}]{milton2018casimir}%
  \BibitemOpen
  \bibfield  {author} {\bibinfo {author} {\bibfnamefont {K.~A.}\ \bibnamefont
  {Milton}}\ and\ \bibinfo {author} {\bibfnamefont {I.}~\bibnamefont
  {Brevik}},\ }\bibfield  {title} {\enquote {\bibinfo {title} {Casimir energies
  for isorefractive or diaphanous balls},}\ }\href@noop {} {\bibfield
  {journal} {\bibinfo  {journal} {Symmetry}\ }\textbf {\bibinfo {volume}
  {10}},\ \bibinfo {pages} {68} (\bibinfo {year} {2018})}\BibitemShut {NoStop}%
\bibitem [{\citenamefont {Milton}\ \emph {et~al.}(1978)\citenamefont {Milton},
  \citenamefont {{DeRaad, Jr}.},\ and\ \citenamefont
  {Schwinger}}]{milton1978casimir}%
  \BibitemOpen
  \bibfield  {author} {\bibinfo {author} {\bibfnamefont {K.~A.}\ \bibnamefont
  {Milton}}, \bibinfo {author} {\bibfnamefont {L.~L.}\ \bibnamefont {{DeRaad,
  Jr}.}}, \ and\ \bibinfo {author} {\bibfnamefont {J.}~\bibnamefont
  {Schwinger}},\ }\bibfield  {title} {\enquote {\bibinfo {title} {Casimir
  self-stress on a perfectly conducting spherical shell},}\ }\href@noop {}
  {\bibfield  {journal} {\bibinfo  {journal} {Ann. Phys. (N. Y.)}\ }\textbf
  {\bibinfo {volume} {115}},\ \bibinfo {pages} {388--403} (\bibinfo {year}
  {1978})}\BibitemShut {NoStop}%
\bibitem [{\citenamefont {Brevik}\ and\ \citenamefont
  {Kolbenstvedt}(1982)}]{brevik1982casimir}%
  \BibitemOpen
  \bibfield  {author} {\bibinfo {author} {\bibfnamefont {I.}~\bibnamefont
  {Brevik}}\ and\ \bibinfo {author} {\bibfnamefont {H.}~\bibnamefont
  {Kolbenstvedt}},\ }\bibfield  {title} {\enquote {\bibinfo {title} {{The
  Casimir effect in a solid ball when $\varepsilon\mu=1$}},}\ }\href@noop {}
  {\bibfield  {journal} {\bibinfo  {journal} {Ann. Phys. (N. Y.)}\ }\textbf
  {\bibinfo {volume} {143}},\ \bibinfo {pages} {179--190} (\bibinfo {year}
  {1982})}\BibitemShut {NoStop}%
\bibitem [{\citenamefont {Avni}\ and\ \citenamefont
  {Leonhardt}(2018)}]{avni2018casimir}%
  \BibitemOpen
  \bibfield  {author} {\bibinfo {author} {\bibfnamefont {Y.}~\bibnamefont
  {Avni}}\ and\ \bibinfo {author} {\bibfnamefont {U.}~\bibnamefont
  {Leonhardt}},\ }\bibfield  {title} {\enquote {\bibinfo {title} {Casimir
  self-stress in a dielectric sphere},}\ }\href@noop {} {\bibfield  {journal}
  {\bibinfo  {journal} {Ann. Phys. (N. Y.)}\ }\textbf {\bibinfo {volume}
  {395}},\ \bibinfo {pages} {326--340} (\bibinfo {year} {2018})}\BibitemShut
  {NoStop}%
\bibitem [{\citenamefont {Efrat}\ and\ \citenamefont
  {Leonhardt}(2021)}]{leonhardt2021}%
  \BibitemOpen
  \bibfield  {author} {\bibinfo {author} {\bibfnamefont {I.~Y.}\ \bibnamefont
  {Efrat}}\ and\ \bibinfo {author} {\bibfnamefont {U.}~\bibnamefont
  {Leonhardt}},\ }\bibfield  {title} {\enquote {\bibinfo {title} {Van der
  {W}aals anomaly},}\ }\href@noop {} {\bibfield  {journal} {\bibinfo  {journal}
  {Phys. Rev. B}\ }\textbf {\bibinfo {volume} {104}},\ \bibinfo {pages}
  {235432} (\bibinfo {year} {2021})}\BibitemShut {NoStop}%
\bibitem [{\citenamefont {Barton}(2001{\natexlab{a}})}]{bartoniv}%
  \BibitemOpen
  \bibfield  {author} {\bibinfo {author} {\bibfnamefont {G}~\bibnamefont
  {Barton}},\ }\bibfield  {title} {\enquote {\bibinfo {title} {Perturbative
  {C}asimir shifts of nondispersive spheres at finite temperature},}\
  }\href@noop {} {\bibfield  {journal} {\bibinfo  {journal} {Phys. Rev. A}\
  }\textbf {\bibinfo {volume} {64}},\ \bibinfo {pages} {032103} (\bibinfo
  {year} {2001}{\natexlab{a}})}\BibitemShut {NoStop}%
\bibitem [{\citenamefont {Milton}(1980)}]{milton1980}%
  \BibitemOpen
  \bibfield  {author} {\bibinfo {author} {\bibfnamefont {K.~A.}\ \bibnamefont
  {Milton}},\ }\bibfield  {title} {\enquote {\bibinfo {title} {Semiclassical
  electron models: {C}asimir self-stress in dielectric and conducting balls},}\
  }\href@noop {} {\bibfield  {journal} {\bibinfo  {journal} {Ann. Phys.
  (N.Y.)}\ }\textbf {\bibinfo {volume} {127}},\ \bibinfo {pages} {49--61}
  (\bibinfo {year} {1980})}\BibitemShut {NoStop}%
\bibitem [{\citenamefont {Milton}\ and\ \citenamefont
  {Ng}(1997)}]{milton1997casimir}%
  \BibitemOpen
  \bibfield  {author} {\bibinfo {author} {\bibfnamefont {K.~A.}\ \bibnamefont
  {Milton}}\ and\ \bibinfo {author} {\bibfnamefont {Y.~J.}\ \bibnamefont
  {Ng}},\ }\bibfield  {title} {\enquote {\bibinfo {title} {{Casimir energy for
  a spherical cavity in a dielectric: Applications to sonoluminescence}},}\
  }\href@noop {} {\bibfield  {journal} {\bibinfo  {journal} {Phys. Rev. E}\
  }\textbf {\bibinfo {volume} {55}},\ \bibinfo {pages} {4207} (\bibinfo {year}
  {1997})}\BibitemShut {NoStop}%
\bibitem [{\citenamefont {Milton}(2001)}]{miltonbook}%
  \BibitemOpen
  \bibfield  {author} {\bibinfo {author} {\bibfnamefont {K.~A.}\ \bibnamefont
  {Milton}},\ }\href@noop {} {\emph {\bibinfo {title} {The Casimir Effect,
  Physical Manifestations of Zero-Point Energy}}}\ (\bibinfo  {publisher}
  {World Scientific},\ \bibinfo {address} {New Jersey},\ \bibinfo {year}
  {2001})\BibitemShut {NoStop}%
\bibitem [{\citenamefont {Schwinger}\ \emph {et~al.}(1998)\citenamefont
  {Schwinger}, \citenamefont {{DeRaad, Jr.}}, \citenamefont {Milton},\ and\
  \citenamefont {Tsai}}]{schwinger2000classical}%
  \BibitemOpen
  \bibfield  {author} {\bibinfo {author} {\bibfnamefont {J.}~\bibnamefont
  {Schwinger}}, \bibinfo {author} {\bibfnamefont {L.~L.}\ \bibnamefont
  {{DeRaad, Jr.}}}, \bibinfo {author} {\bibfnamefont {K.~A.}\ \bibnamefont
  {Milton}}, \ and\ \bibinfo {author} {\bibfnamefont {W.-y.}\ \bibnamefont
  {Tsai}},\ }\href@noop {} {\emph {\bibinfo {title} {Classical
  Electrodynamics}}}\ (\bibinfo  {publisher} {Perseus/Westview and Taylor and
  Francis},\ \bibinfo {year} {1998})\BibitemShut {NoStop}%
\bibitem [{\citenamefont {Parashar}\ \emph {et~al.}(2017)\citenamefont
  {Parashar}, \citenamefont {Milton}, \citenamefont {Shajesh},\ and\
  \citenamefont {Brevik}}]{parashar2017electromagnetic}%
  \BibitemOpen
  \bibfield  {author} {\bibinfo {author} {\bibfnamefont {P.}~\bibnamefont
  {Parashar}}, \bibinfo {author} {\bibfnamefont {K.~A.}\ \bibnamefont
  {Milton}}, \bibinfo {author} {\bibfnamefont {K.~V.}\ \bibnamefont {Shajesh}},
  \ and\ \bibinfo {author} {\bibfnamefont {I.}~\bibnamefont {Brevik}},\
  }\bibfield  {title} {\enquote {\bibinfo {title} {Electromagnetic
  $\delta$-function sphere},}\ }\href@noop {} {\bibfield  {journal} {\bibinfo
  {journal} {Phys. Rev. D}\ }\textbf {\bibinfo {volume} {96}},\ \bibinfo
  {pages} {085010} (\bibinfo {year} {2017})}\BibitemShut {NoStop}%
\bibitem [{\citenamefont {Milton}(2011)}]{milton2011local}%
  \BibitemOpen
  \bibfield  {author} {\bibinfo {author} {\bibfnamefont {K.~A.}\ \bibnamefont
  {Milton}},\ }\bibfield  {title} {\enquote {\bibinfo {title} {{Local and
  global Casimir energies: Divergences, renormalization, and the coupling to
  gravity}},}\ }in\ \href@noop {} {\emph {\bibinfo {booktitle} {Casimir
  Physics}}}\ (\bibinfo  {publisher} {Springer},\ \bibinfo {year} {2011})\ pp.\
  \bibinfo {pages} {39--95}\BibitemShut {NoStop}%
\bibitem [{\citenamefont {Balian}\ and\ \citenamefont
  {Duplantier}(1978)}]{balian1978electromagnetic}%
  \BibitemOpen
  \bibfield  {author} {\bibinfo {author} {\bibfnamefont {R.}~\bibnamefont
  {Balian}}\ and\ \bibinfo {author} {\bibfnamefont {B.}~\bibnamefont
  {Duplantier}},\ }\bibfield  {title} {\enquote {\bibinfo {title}
  {{Electromagnetic waves near perfect conductors. II. Casimir effect}},}\
  }\href@noop {} {\bibfield  {journal} {\bibinfo  {journal} {Ann. Phys. (N.
  Y.)}\ }\textbf {\bibinfo {volume} {112}},\ \bibinfo {pages} {165--208}
  (\bibinfo {year} {1978})}\BibitemShut {NoStop}%
\bibitem [{\citenamefont {Milton}\ \emph {et~al.}(2013)\citenamefont {Milton},
  \citenamefont {Kheirandish}, \citenamefont {Parashar}, \citenamefont {Abalo},
  \citenamefont {Fulling}, \citenamefont {Bouas}, \citenamefont {Carter},\ and\
  \citenamefont {Kirsten}}]{torque1}%
  \BibitemOpen
  \bibfield  {author} {\bibinfo {author} {\bibfnamefont {K.~A.}\ \bibnamefont
  {Milton}}, \bibinfo {author} {\bibfnamefont {F.}~\bibnamefont {Kheirandish}},
  \bibinfo {author} {\bibfnamefont {P.}~\bibnamefont {Parashar}}, \bibinfo
  {author} {\bibfnamefont {E.~K.}\ \bibnamefont {Abalo}}, \bibinfo {author}
  {\bibfnamefont {S.~A.}\ \bibnamefont {Fulling}}, \bibinfo {author}
  {\bibfnamefont {J.~D.}\ \bibnamefont {Bouas}}, \bibinfo {author}
  {\bibfnamefont {H.}~\bibnamefont {Carter}}, \ and\ \bibinfo {author}
  {\bibfnamefont {K.}~\bibnamefont {Kirsten}},\ }\bibfield  {title} {\enquote
  {\bibinfo {title} {Investigation of the torque anomaly in an annular sector.
  {I}. {Global} calculations, scalar case},}\ }\href@noop {} {\bibfield
  {journal} {\bibinfo  {journal} {Phys. Rev. D}\ }\textbf {\bibinfo {volume}
  {88}},\ \bibinfo {pages} {025039} (\bibinfo {year} {2013})}\BibitemShut
  {NoStop}%
\bibitem [{\citenamefont {Nesterenko}\ \emph {et~al.}(2001)\citenamefont
  {Nesterenko}, \citenamefont {Lambiase},\ and\ \citenamefont
  {Scarpetta}}]{nesterenko2001casimir}%
  \BibitemOpen
  \bibfield  {author} {\bibinfo {author} {\bibfnamefont {V.~V.}\ \bibnamefont
  {Nesterenko}}, \bibinfo {author} {\bibfnamefont {G.}~\bibnamefont
  {Lambiase}}, \ and\ \bibinfo {author} {\bibfnamefont {G.}~\bibnamefont
  {Scarpetta}},\ }\bibfield  {title} {\enquote {\bibinfo {title} {Casimir
  effect for a dilute dielectric ball at finite temperature},}\ }\href@noop {}
  {\bibfield  {journal} {\bibinfo  {journal} {Phys. Rev. D}\ }\textbf {\bibinfo
  {volume} {64}},\ \bibinfo {pages} {025013} (\bibinfo {year}
  {2001})}\BibitemShut {NoStop}%
\bibitem [{\citenamefont
  {Barton}(2001{\natexlab{b}})}]{barton2001perturbative}%
  \BibitemOpen
  \bibfield  {author} {\bibinfo {author} {\bibfnamefont {G.}~\bibnamefont
  {Barton}},\ }\bibfield  {title} {\enquote {\bibinfo {title} {{Perturbative
  Casimir shifts of dispersive spheres at finite temperature}},}\ }\href@noop
  {} {\bibfield  {journal} {\bibinfo  {journal} {J. Phys. A: Math. Theor.}\
  }\textbf {\bibinfo {volume} {34}},\ \bibinfo {pages} {5781} (\bibinfo {year}
  {2001}{\natexlab{b}})}\BibitemShut {NoStop}%
\bibitem [{\citenamefont {Milton}\ and\ \citenamefont
  {Ng}(1998)}]{milton1998vdw}%
  \BibitemOpen
  \bibfield  {author} {\bibinfo {author} {\bibfnamefont {K.~A.}\ \bibnamefont
  {Milton}}\ and\ \bibinfo {author} {\bibfnamefont {Y.~J.}\ \bibnamefont
  {Ng}},\ }\bibfield  {title} {\enquote {\bibinfo {title} {Observability of the
  bulk {C}asimir effect: {C}an the dynamical {C}asimir effect be relevant to
  sonoluminescence?}}\ }\href@noop {} {\bibfield  {journal} {\bibinfo
  {journal} {Phys. Rev. E}\ }\textbf {\bibinfo {volume} {57}},\ \bibinfo
  {pages} {5504} (\bibinfo {year} {1998})}\BibitemShut {NoStop}%
\bibitem [{\citenamefont {Brevik}\ \emph {et~al.}(1999)\citenamefont {Brevik},
  \citenamefont {Marachevsky},\ and\ \citenamefont
  {Milton}}]{brevik1999identity}%
  \BibitemOpen
  \bibfield  {author} {\bibinfo {author} {\bibfnamefont {I.}~\bibnamefont
  {Brevik}}, \bibinfo {author} {\bibfnamefont {V.~N.}\ \bibnamefont
  {Marachevsky}}, \ and\ \bibinfo {author} {\bibfnamefont {K.~A.}\ \bibnamefont
  {Milton}},\ }\bibfield  {title} {\enquote {\bibinfo {title} {{Identity of the
  van der Waals force and the Casimir effect and the irrelevance of these
  phenomena to sonoluminescence}},}\ }\href@noop {} {\bibfield  {journal}
  {\bibinfo  {journal} {Phys. Rev. Lett.}\ }\textbf {\bibinfo {volume} {82}},\
  \bibinfo {pages} {3948} (\bibinfo {year} {1999})}\BibitemShut {NoStop}%
\bibitem [{\citenamefont {Dzyaloshinskii}\ \emph {et~al.}(1961)\citenamefont
  {Dzyaloshinskii}, \citenamefont {Lifshitz},\ and\ \citenamefont
  {Pitaevskii}}]{dlp}%
  \BibitemOpen
  \bibfield  {author} {\bibinfo {author} {\bibfnamefont {I.~E}\ \bibnamefont
  {Dzyaloshinskii}}, \bibinfo {author} {\bibfnamefont {E.~M}\ \bibnamefont
  {Lifshitz}}, \ and\ \bibinfo {author} {\bibfnamefont {L.~P.}\ \bibnamefont
  {Pitaevskii}},\ }\bibfield  {title} {\enquote {\bibinfo {title} {The general
  theory of van der {W}aals forces},}\ }\href@noop {} {\bibfield  {journal}
  {\bibinfo  {journal} {Adv. Phys.}\ }\textbf {\bibinfo {volume} {10}},\
  \bibinfo {pages} {165--209} (\bibinfo {year} {1961})}\BibitemShut {NoStop}%
\bibitem [{\citenamefont {Christensen}(1976)}]{christensen}%
  \BibitemOpen
  \bibfield  {author} {\bibinfo {author} {\bibfnamefont {S.~M.}\ \bibnamefont
  {Christensen}},\ }\bibfield  {title} {\enquote {\bibinfo {title} {Vacuum
  expectation value of the stress tensor in an arbitrary curved background: The
  covariant point-separation method},}\ }\href@noop {} {\bibfield  {journal}
  {\bibinfo  {journal} {Phys. Rev. D}\ }\textbf {\bibinfo {volume} {14}},\
  \bibinfo {pages} {2490--2501} (\bibinfo {year} {1976})}\BibitemShut {NoStop}%
\bibitem [{\citenamefont {Fraser-McKelvie}\ \emph {et~al.}(2011)\citenamefont
  {Fraser-McKelvie}, \citenamefont {Pimbblet},\ and\ \citenamefont
  {Lazendic}}]{density}%
  \BibitemOpen
  \bibfield  {author} {\bibinfo {author} {\bibfnamefont {A}~\bibnamefont
  {Fraser-McKelvie}}, \bibinfo {author} {\bibfnamefont {K.~A.}\ \bibnamefont
  {Pimbblet}}, \ and\ \bibinfo {author} {\bibfnamefont {J.~S.}\ \bibnamefont
  {Lazendic}},\ }\bibfield  {title} {\enquote {\bibinfo {title} {An estimate of
  the electron density in filaments of galaxies at $z\sim 0.1$},}\ }\href@noop
  {} {\bibfield  {journal} {\bibinfo  {journal} {Mon. Not. R. Astron. Soc.}\
  }\textbf {\bibinfo {volume} {415}},\ \bibinfo {pages} {1961} (\bibinfo {year}
  {2011})}\BibitemShut {NoStop}%
\bibitem [{\citenamefont {Borwein}\ and\ \citenamefont
  {Borwein}(1987)}]{borwein2}%
  \BibitemOpen
  \bibfield  {author} {\bibinfo {author} {\bibfnamefont {J.~M}\ \bibnamefont
  {Borwein}}\ and\ \bibinfo {author} {\bibfnamefont {P.~B.}\ \bibnamefont
  {Borwein}},\ }\href@noop {} {\emph {\bibinfo {title} {Pi and the AGM: A Study
  in Analytic Number Theory and Computational Complexity}}},\ Canadian
  Mathematical Society Series of Monographs and Advanced Texts, Volume 4\
  (\bibinfo  {publisher} {John Wiley and Sons},\ \bibinfo {address} {New
  York},\ \bibinfo {year} {1987})\BibitemShut {NoStop}%
\bibitem [{\citenamefont {Breuer}\ and\ \citenamefont
  {Petruccione}(2002)}]{breuer}%
  \BibitemOpen
  \bibfield  {author} {\bibinfo {author} {\bibfnamefont {H.~P.}\ \bibnamefont
  {Breuer}}\ and\ \bibinfo {author} {\bibfnamefont {F.}~\bibnamefont
  {Petruccione}},\ }\href@noop {} {\emph {\bibinfo {title} {The Theory of Open
  Quantum Systens}}}\ (\bibinfo  {publisher} {Oxford University Press},\
  \bibinfo {address} {New York},\ \bibinfo {year} {2002})\BibitemShut {NoStop}%
\bibitem [{\citenamefont {Lifshitz}(1956)}]{lifshitz}%
  \BibitemOpen
  \bibfield  {author} {\bibinfo {author} {\bibfnamefont {E.~M.}\ \bibnamefont
  {Lifshitz}},\ }\bibfield  {title} {\enquote {\bibinfo {title} {The theory of
  molecular attractive forces between solids},}\ }\href@noop {} {\bibfield
  {journal} {\bibinfo  {journal} {Sov. Phys. JETP}\ }\textbf {\bibinfo {volume}
  {2}},\ \bibinfo {pages} {73--83} (\bibinfo {year} {1956})}\BibitemShut
  {NoStop}%
\bibitem [{\citenamefont {Casimir}\ and\ \citenamefont
  {Polder}(1948)}]{casimirpolder}%
  \BibitemOpen
  \bibfield  {author} {\bibinfo {author} {\bibfnamefont {H.~B.~G.}\
  \bibnamefont {Casimir}}\ and\ \bibinfo {author} {\bibfnamefont
  {D.}~\bibnamefont {Polder}},\ }\bibfield  {title} {\enquote {\bibinfo {title}
  {The influence of retardation on the {L}ondon-van der {W}aals forces},}\
  }\href@noop {} {\bibfield  {journal} {\bibinfo  {journal} {Phys. Rev.}\
  }\textbf {\bibinfo {volume} {73}},\ \bibinfo {pages} {360--372} (\bibinfo
  {year} {1948})}\BibitemShut {NoStop}%
\bibitem [{\citenamefont {Schwinger}\ \emph {et~al.}(1978)\citenamefont
  {Schwinger}, \citenamefont {DeRaad},\ and\ \citenamefont {Milton}}]{sdm1978}%
  \BibitemOpen
  \bibfield  {author} {\bibinfo {author} {\bibfnamefont {J.}~\bibnamefont
  {Schwinger}}, \bibinfo {author} {\bibfnamefont {L.~L.}\ \bibnamefont
  {DeRaad}, \bibfnamefont {Jr.}}, \ and\ \bibinfo {author} {\bibfnamefont
  {K.~A.}\ \bibnamefont {Milton}},\ }\bibfield  {title} {\enquote {\bibinfo
  {title} {Casimir effect in dielectrics},}\ }\href@noop {} {\bibfield
  {journal} {\bibinfo  {journal} {Ann. Phys. (N.Y)}\ }\textbf {\bibinfo
  {volume} {115}},\ \bibinfo {pages} {1--23} (\bibinfo {year}
  {1978})}\BibitemShut {NoStop}%
\end{thebibliography}%

\end{document}